\documentclass[aps, prd, amsmath, amssymb, floats, floatfix, superscriptaddress, nofootinbib, twocolumn, showpacs,reprint]{revtex4-1}

\usepackage{graphicx}
\usepackage{amsmath}
\usepackage{amssymb}
\usepackage{amsfonts}
\usepackage{times}
\usepackage{xspace} 
\usepackage[usenames]{color}
\usepackage{dcolumn}
\usepackage{bm}
\usepackage{mathrsfs}
\usepackage{tikz}
\usepackage{appendix}
\usepackage{breqn}
\usepackage[normalem]{ulem}

\usepackage[tracking=true]{microtype}
\SetTracking{}{500}
\SetTracking{encoding={*}, shape=sc}{40}

\newcommand{\cs}{c_\sigma}
\newcommand{\ca}{c_a}

\makeatletter
\let\cat@comma@active\@empty
\makeatother

\begin{document}

\title{Slowly rotating black holes in Einstein-\ae ther theory}

\author{Enrico Barausse}
\affiliation{CNRS, UMR 7095, Institut d'Astrophysique de Paris, 98bis Bd Arago, 75014 Paris, France}
\affiliation{Sorbonne Universit\'es, UPMC Univ Paris 06, UMR 7095, 98bis Bd Arago, 75014 Paris, France}

\author{Thomas P.~Sotiriou}
\affiliation{School of Mathematical Sciences \& School of Physics and Astronomy, University of Nottingham, University Park, Nottingham, NG7 2RD, UK}

\author{Ian Vega}
\affiliation{National Institute of Physics, University of the Philippines, Diliman, Quezon City 1101, Philippines}
\affiliation{SISSA, Via Bonomea 265, 34136, Trieste, Italy and INFN, Sezione di Trieste, Italy}

\begin{abstract}
We study slowly rotating, asymptotically flat black holes in Einstein-\ae ther theory and show that solutions that are free from naked finite area singularities form a two-parameter family. These parameters can be thought of as the mass and angular momentum of the black hole, while there are no independent \ae ther charges. We also show that the \ae ther has non-vanishing vorticity throughout the spacetime, as a result of which there is no hypersurface that resembles the universal horizon found in static, spherically symmetric solutions. 
Moreover, for experimentally viable choices of the coupling constants, the frame-dragging potential of our solutions only shows percent-level deviations
from the corresponding quantities in General Relativity and Ho\v rava gravity.
Finally, we uncover and discuss several subtleties in the correspondence between Einstein-\ae ther theory and Ho\v rava gravity solutions in the $c_\omega\to \infty$ limit.
\end{abstract}

\maketitle

\section{Introduction}

Einstein-\ae ther  theory (\ae-theory) \cite{Jacobson:2004ts} is essentially General Relativity (GR) coupled with a unit-norm, timelike vector field, $u^\mu$, usually referred to as the ``\ae ther''. The unit-norm timelike constraint on the \ae ther forces it to be ever-present, even in the local frame, thus selecting a preferred time direction and violating local Lorentz symmetry. The action for the \ae ther contains all possible terms that are quadratic in the first derivatives of $u^\mu$ (up to total divergences). Hence, \ae-theory can be considered as an effective description of Lorentz symmetry breaking in the gravity sector. Indeed, it has been extensively used in order to obtain quantitative constraints on Lorentz-violating gravity (see Ref.~\cite{Jacobson:2008aj} for a review on \ae-theory).  
Additionally, violations of Lorentz symmetry in the gravitational sector have been used to construct modified-gravity theories that account for Dark-Matter phenomenology without any actual Dark Matter~\cite{Bekenstein:2004ne,Zlosnik:2006sb,Zlosnik:2006zu,Blanchet:2011wv,Bonetti:2015oda}.

The action for  \ae-theory can be written as~\cite{Jacobson:2013xta}
\begin{multline}\label{aeaction}
	S_{\ae}=-\frac{1}{16\pi G_{\ae}}\int \Big(R + \frac{1}{3}c_\theta \theta^2 + c_\sigma \sigma_{\mu\nu} \sigma^{\mu\nu} \\
	+ c_\omega \omega_{\mu\nu} \omega^{\mu\nu} + c_a a_\mu a^\mu\Big) \sqrt{-g} d^4x\,
\end{multline}
where $c_\theta$, $c_a$, $c_\sigma$ and $c_\omega$ are dimensionless coupling constants, while $\theta$, $a_\mu$, $\sigma_{\mu\nu}$ and $\omega_{\mu\nu}$ are respectively the expansion, acceleration, shear and vorticity
of the congruence defined by the vector field $u^\mu$:
\begin{gather}
\theta=\nabla_\mu u^\mu\,,\\
a^\mu= u^\nu\nabla_\nu u^\mu\,,\\ 
\sigma_{\mu\nu}=\nabla_{(\mu}u_{\nu)}-a_{(\nu}u_{\mu)}+\frac13\theta h_{\mu\nu}\,, \\
\omega_{\mu\nu}=\nabla_{[\mu}u_{\nu]}- a_{[\nu}u_{\mu]}=\partial_{[\mu}u_{\nu]}-a_{[\nu}u_{\mu]}\,. \label{eqn:vort}
\end{gather}
where $h_{\mu\nu}=g_{\mu\nu}-u_\mu u_\nu$ is the projector orthogonal to the \ae ther,
and we are assuming a metric signature $(+,-,-,-)$ and setting $c=1$. (We will stick to these conventions
throughout this paper.) The constraint $g_{\mu\nu} u^\mu u^\nu =1$ that forces the \ae ther to be unit-norm and timelike can be imposed either by a Lagrange multiplier, or implicitly by restricting the \ae ther variations to be normal to the \ae ther when applying variational principles. 
The bare gravitational constant $G_{\ae}$ is related to the gravitational constant $G$
(as measured by torsion-pendulum experiments) by $G=2G_{\ae}/(2-c_a)$~\cite{Carroll:2004ai}.
Note that we will adopt units where $G=1$ throughout this paper. When added to the theory, matter is assumed to couple minimally to the metric $g_{\mu\nu}$ and not directly to the \ae ther. This guarantees that the weak equivalence principle is satisfied. 

A perturbative analysis over a Minkowski background reveals that 
\ae-theory contains not only spin-2 gravitons (like GR), but also spin-1 and spin-0 polarizations~\cite{Jacobson:2004ts}. The flat-space propagation
speeds $s_2$, $s_1$ and $s_0$ of these graviton modes depend on the coupling constants 
introduced above:
\begin{align}
	s_0^2 &= \frac{(c_\theta + 2c_\sigma)(1-c_a/2)}{3c_a(1-c_\sigma)(1+c_\theta/2)} \label{eqn:s0mode}\,,\\
	s_1^2 &= \frac{c_\sigma + c_\omega (1-c_\sigma)}{2c_a(1-c_\sigma)} \label{eqn:s1mode}\,,\\
	s_2^2 &= \frac{1}{1-c_\sigma}\,.
\end{align}

In order to ensure classical and quantum stability (no gradient instabilities and no ghosts), it is 
necessary and sufficient that $s_i^2>0$ (with $i=1,2,3$)~\cite{Jacobson:2004ts,Jacobson:2008aj}. Furthermore, constraints from cosmic-ray observations require that the speeds of massless excitations be luminal or superluminal \cite{Elliott:2005va}. If this were not the case, the energy of cosmic rays (which travel at relativistic speeds) would dissipate 
into subluminal massless modes via a \v Cerenkov-like process, and it would not be possible to account for the high cosmic-ray energies that we actually observe. Additional constraints then come from requiring agreement with solar-system~\cite{Foster:2005dk,Jacobson:2008aj} and cosmological~\cite{Zuntz:2008zz} tests, and most of all, with isolated- and binary-pulsar observations~\cite{Yagi:2013qpa,Yagi:2013ava}. As a result, the dimensionless couplings  $c_\theta$, $c_a$, $c_\sigma$ and $c_\omega$ are required to be close to the GR limit  $c_\theta=c_a=c_\sigma=c_\omega=0$, i.e. $|c_\theta|,|c_a|,|c_\sigma|,|c_\omega|\lesssim \mbox{a few }\times 0.01$~\cite{Yagi:2013qpa,Yagi:2013ava}. Since these coupling constants have to be ``small'', for most purposes one can expand the theory's dynamics perturbatively in the couplings. We will indeed adopt this ``small-coupling limit'' in some of the calculations of this paper.

The presence of a preferred frame violating Lorentz-invariance mitigates the causality concerns that one would have in GR regarding superluminal motion. However, the existence of superluminal excitations ought to be relevant for black holes. In GR, stationary black holes are defined by their event horizons, which can be understood as null hypersurfaces of the metric 
(in our case $g_{\mu\nu}$) to which photons (and more generally matter) couple minimally. These hypersurfaces will act as one-way causal boundaries for luminal or subluminal excitations. However, superluminal excitations could penetrate them in both directions. 

The resolution to this apparent conundrum lies in the fact that null congruences with respect to $g_{\mu\nu}$ do not actually determine causality, if Lorentz symmetry is violated. In fact, causality in \ae-theory should be dictated by the characteristics of its field equations \cite{Eling:2006ec}.
These are  determined by looking at high-frequency solutions to the linearized field equations, i.e. the characteristics
are essentially the null cones along which the different excitations propagate in the eikonal limit, and for each 
spin-$i$ mode they can be shown~\cite{Eling:2006ec}
to be null hypersurfaces of the effective metric 
\begin{equation}
	g^{(i)}_{\mu\nu} = g_{\mu\nu} + (s_i^2 -1)u_\mu u_\nu
	\label{eqn:hors}
\end{equation}
where $s_i$ is the mode's flat-spacetime propagation speed (with respect to the \ae ther rest frame).

Based on the above, one expects a black hole to possess multiple horizons, i.e. at least one for each excitation traveling at a given speed. The relative spacetime location of these horizons will depend on the relative speeds of the different excitations, with the ``slowest'' excitation having the outermost horizon. Indeed, this intuitive picture agrees completely with the outcome of studies of static, spherically symmetric black holes in \ae-theory \cite{Eling:2006ec,Barausse-etal-PRD:2011}. Remarkably though, those black holes exhibit another crucial feature \cite{Barausse-etal-PRD:2011}. The \ae ther, which is hypersurface-orthogonal due to the assumption of spherical symmetry, becomes normal to one or more constant-radius hypersurfaces that lie inside the Killing horizon of $g_{\mu\nu}$. What makes this feature remarkable is that the \ae ther, by definition, determines the preferred time direction, which then implies that any hypersurface to which it is normal can only be crossed in one direction, else one would be traveling toward the past. These hypersurfaces are particularly relevant for the causal structure, because they do not distinguish between the speeds or any other characteristic of an excitation, and act as causal boundaries for any propagating mode on the mere assumption that motion is future directed. Because of this property, these hypersurfaces were called {\em universal horizons} \cite{Barausse-etal-PRD:2011,blas-sibiryakov-prd:2011}. 

The relevance of universal horizons to the causal structure of black holes in \ae-theory is likely to be limited, as they are cloaked by the more conventional excitation-specific horizons. However, an ultraviolet completion of \ae-theory is likely to involve higher-order dispersion relations, because once Lorentz symmetry is abandoned there is no particular reason to expect the dispersion relation to remain linear. Indeed, it has been shown in Ref.~\cite{Jacobson:2013xta} that an action that is formally the same as in \ae-theory, but in which the \ae ther is forced to be hypersurface-orthogonal {\em a priori} (before the variation), corresponds to the low-energy limit of Ho\v rava gravity \cite{Horava:2009uw}. The latter is a power-counting renormalizable gravity theory with a preferred foliation (as opposed to just a preferred frame) and higher-order dispersion relations (see Refs.~\cite{Sotiriou:2010wn,Blas:2014aca,Berti:2015itd} for reviews).  Given the correspondence between the two theories and the fact that in spherical symmetry vectors are hypersurface-orthogonal, it is clear that spherical black-hole solutions of \ae-theory are also solutions of Ho\v rava gravity (the reverse is not as straightforward, 
but holds true as well for static, spherically symmetric and asymptotically flat black holes \cite{Barausse:2013nwa,Barausse:2012ny,Blas:2010hb}). Indeed, Ref.~\cite{Barausse-etal-PRD:2011} considered both theories, while universal horizons have been found in the small-coupling limit of Ho\v rava gravity in Ref.~\cite{blas-sibiryakov-prd:2011}.

We will discuss the main characteristics of Ho\v rava gravity and the relation between the two theories in more detail in a forthcoming section. What is worth mentioning here is that once the higher-order terms in the dispersion relation are taken into account, perturbations with sufficiently short wavelength can travel arbitrarily fast, making 
the universal horizon the only relevant causal boundary. This makes universal horizons particularly interesting in Ho\v rava gravity, and potentially in ultraviolet completions of \ae-theory. Without them, the notion of a black hole in these theories would be merely a low-energy artifact.

So far we have based our discussions on results that assume staticity and spherical symmetry. Recently, the concept of a universal horizon in theories with a preferred foliation has been discussed in detail and defined rigorously without any reference to specific symmetries \cite{Bhattacharyya:2015gwa}. However, actual solutions beyond spherical symmetry are sparse. Stationary, axisymmetric solutions have been considered in Ref.~\cite{Sotiriou:2014gna} in special sectors of Ho\^{r}ava gravity in three dimensions, and it has been shown that the existence of universal horizons is a rather generic feature in these black-hole solutions. Remarkably, universal horizons that lie beyond cosmological de Sitter horizons in solutions with suitable asymptotics have been discovered. However, solutions without universal horizons have also been found. In four dimensions, it has been shown in Ref.~\cite{Barausse:2012qh} that slowly rotating black holes in the infrared limit of Ho\v rava gravity continue to possess a universal horizon, whereas in \ae-theory the \ae ther ceases to be globally hypersurface-orthogonal once rotation is taken into account. That is, even though the two theories share spherically symmetric solutions, they do not share rotating ones. This is an indication that rotating solutions in \ae-theory may not possess universal horizons, but it is far from a definitive proof. The potential loophole is for the \ae ther to be orthogonal to a \emph{specific} hypersurface without being globally hypersurface-orthogonal. This special hypersurface could then potentially play the role of a universal horizon. Exploring this possibility is one of the aims of this paper.

More generally, in the following we study slowly rotating black hole solutions in \ae-theory. We  build on the results of Ref.~\cite{Barausse:2013nwa}, which has shown that in the slowly rotating limit the \ae-theory equation can be reduced to a pair of coupled ordinary differential equations (ODEs). We consider in detail the structure of the equations, which reveals the need to impose a regularity condition if naked finite-area singularities are to be avoided. This condition, together with asymptotic flatness, pins down the number of independent parameters in the solution to two, the mass and the angular momentum. Hence, we show that the \ae ther cannot carry any independent hair. We then 
solve the equations explicitly to generate slowly rotating solutions, both 
in the small-coupling approximation (in which the backreaction of the \ae ther's rotation on the spherically symmetric background metric is neglected), and 
in the general case.
We show that in both cases the solutions do not possess a universal horizon, unless the coupling constants take specific values for which
the solutions reduce to those of Ho\v rava gravity. Finally, we discuss in some detail this latter point, i.e. the limit of \ae-theory solutions to Ho\v rava gravity ones, and we highlight a subtlety that had passed unnoticed so far. 

\section{Methodology}
\subsection{The field equations in the slow-rotation limit}\label{field_eqs}
The metric describing a slowly rotating body 
is given by the well-known Hartle-Thorne ansatz \cite{HartleThorne68}
\begin{equation}
	\begin{split}
ds^2 = f(r) dt^2 &-\frac{B(r)^2}{f(r)}dr^2 - r^2 (d\theta^2 + \sin^2\theta d\phi^2)\\ &+ \epsilon r^2 \sin^2\theta \Omega(r,\theta) dtd\phi + O(\epsilon^2)\,,\label{metric}
	\end{split}
\end{equation}
where $f(r)$ and $B(r)$ characterize the ``seed'' static, spherically symmetric solutions when the ``frame dragging'' $\Omega(r,\theta)$ is set to zero,
and $\epsilon$ is a perturbative ``slow-rotation'' parameter. Using arguments similar to those used by Hartle and Thorne for the metric ansatz, one can show that in the slow-rotation limit the \ae ther field can be described by~\cite{Barausse:2013nwa} 
\begin{equation}
	\begin{split}
u_\alpha dx^\alpha = &\frac{1+f(r)A(r)^2}{2A(r)}dt +\frac{B(r)}{2A(r)}\left[\frac{1}{f(r)}-A(r)^2\right]dr \\ &+ \epsilon \left[\frac{1+f(r)A(r)^2}{2A(r)}\right]\lambda(r,\theta) \sin^2\theta d\phi + O(\epsilon^2),
\label{eqn:aether}
	\end{split}
\end{equation}
where $A(r)$ is a potential characterizing the static, spherically symmetric solution, while $\lambda(r,\theta)$ is related to
the \ae ther's angular momentum per unit energy by $u_\phi/u_t = \lambda(r,\theta)\sin^2\theta$.

It has been shown in Ref.~\cite{Barausse:2013nwa} that $\Omega$ and $\lambda$ have to be independent of $\theta$, {\em i.e.}~$\Omega(r,\theta)=\Omega(r)$ and $\lambda(r,\theta)=\lambda(r)$, if the solutions are to be regular at the poles and the \ae ther is to be asymptotically at rest at spatial infinity, where the metric becomes asymptotically flat~\footnote{In the following,
we will utilize the expression ``asymptotic flatness''
to denote, for brevity's sake, two conditions that are to be satisfied at the same time, i.e. that the metric approaches the Minkowski one at large radii, and
that the \ae ther asymptotically aligns with the timelike Killing vector (thus being asymptotically at rest).}. Under this assumption and by introducing the \ae ther's angular velocity
\begin{equation}
\psi(r) = \frac{u^\phi}{u^t}= \frac{1}{2} \Omega(r) - \frac{f(r)\lambda(r)}{r^2}\,,
\label{eqn:psidef}
\end{equation}
 the field equations at order ${\cal O}(\epsilon)$ reduce to the following coupled, homogeneous linear ordinary differential equations for $\psi(r)$ and $\lambda(r)$~\cite{Barausse:2013nwa}:
\begin{equation}\label{eeq1}
d_1(r) \psi'(r)+d_2(r) \psi''(r)+d_3(r) \lambda'(r)+d_4(r) \lambda''(r)=0
\end{equation}
 and
\begin{equation}\label{eeq2}
b_1(r) \psi'(r)+b_2(r) \psi''(r)+b_3(r) \lambda'(r)+b_4(r) \lambda''(r)+\frac{\lambda(r)}{r^4}=0\,,
\end{equation}
where the prime denotes differentiation with respect to $r$, and the coefficients $\{d_i,b_i\}$ are functions of $\{f,B,A\}$ given explicitly in Appendix \ref{app:coeffs}.
For the purposes of this paper, however, it is convenient to adopt $\Omega(r)$ and $\Lambda=\lambda (1+fA^2)/(2A)$
as our variables. The field equations therefore become
\begin{subequations}
\label{eqns:new}
\begin{align}
\Omega'' &= \frac{1}{S}\left(p_1\Omega' + p_2\Lambda'+ \frac{p_3}{U} \Lambda\right), \label{eqn:oeqnFull}\\ 
\Lambda'' &= \frac{1}{S}\left(q_1\Omega' + q_2\Lambda' + \frac{q_3}{U} \Lambda\right), \label{eqn:ueqnFull}
\end{align}
\end{subequations}
where 
\begin{align}
U(r) \equiv & \,\,1+ fA^2\propto u_t\,,
	\label{eqn:sing1}\\
S(r) \equiv & \,\,(s_1^2-1)(1+fA^2)^2 + 4fA^2 \notag\\&\propto g^{(1)}_{tt}= g_{tt} + (s_1^2 -1)u_t^2\,.
	\label{eqn:sing2}
\end{align}
The functions $U$ and $S$ vanish at the universal horizon and spin-1 horizon, respectively. The coefficients $\{p_i, q_i\}$ are functions of $\{f,B,A\}$ and their explicit expressions are given in Appendix \ref{pandq}.
These coefficients are well behaved everywhere as long as $\{f,B,A\}$ are regular. Therefore, the locations of the universal and spin-1 horizons are the only possible singularities of Eqs.~(\ref{eqns:new}). 

\subsection{Boundary conditions}
\label{bound}

Our goal is to find asymptotically flat solutions to Eqs.~\eqref{eqns:new} that are regular everywhere, except possibly at their center. However, as we have seen above, Eqs.~\eqref{eqns:new} exhibit apparent singularities on the spin-1  and universal horizons of the spherically symmetric ``seed'' solution, where $S=0$ and $U=0$ respectively. Let us first address  the behavior of the solutions on the universal horizon. If the solutions are to be regular there,
the terms $\Lambda\, p_3/U$ and $\Lambda\, q_3/U$
should remain finite when $U=0$. There are therefore two distinct options: 
either $\Lambda={\cal O}(U)$, or $p_3={\cal O}(U)$ and $q_3={\cal O}(U)$. Following the first option, one could consider imposing $\Lambda=0$ on the universal horizon of the spherical ``seed'' solution as an additional condition. Such a condition would inevitably reduce the number of independent parameters characterizing the solution. It turns out that one need not resort to this. 
Considering indeed the other option, by using the nonlinear equations for $\{f,B,A\}$~\cite{Barausse-etal-PRD:2011} we have verified that both $q_3/U$ and $p_3/U$ are actually regular at $U=0$. Indeed, eliminating the highest-order derivatives of $\{f,B,A\}$ by using the background field equations, namely Eqs.~(36) - (38) of Ref.~\cite{Barausse-etal-PRD:2011}, is enough to show that $q_3/U$ is regular when $U=0$.  
To show that $p_3/U$ is regular, one also has to use one of the constraint equations for the background, Eqs.~(35) of Ref.~\cite{Barausse-etal-PRD:2011}.
Hence, we conclude that no additional regularity condition is required on the universal horizon.
Since the demonstration sketched above involves cumbersome equations and is in general not very instructive, in the following we will only 
present it explicitly in the small coupling limit (c.f. Sec.~\ref{sec:small_couplings}), and for two special choices of couplings for which exact spherically symmetric solutions are known (c.f. Secs.~\ref{sec:mattingly_solns1} and \ref{sec:cazero}).

We now move on to the behavior of the solutions on the spin-1 horizon, where $S=0$. Generic solutions will indeed be singular there. To see this, we can look at the curvature scalar $R_{t\phi} (\equiv R_{\alpha\beta}t^\alpha \phi^\beta)$, which can be verified to depend on $\Omega'(r)$ and $\Omega''(r)$, and note that barring fine-tuning of $\Omega'$, $\Lambda$ and $\Lambda'$, 
Eq. (\ref{eqn:oeqnFull}) implies that $\Omega''$ (and thus $R_{t\phi}$) will generically diverge. 
To ensure that this does not occur, we need to impose 
regularity of Eq. (\ref{eqn:oeqnFull}) at the spin-1 horizon $r=r_s$, i.e.
\begin{equation}
	p_1(r_s)\Omega'(r_s) + p_2(r_s) \Lambda'(r_s) = -\frac{p_3(r_s)}{U(r_s)} \Lambda(r_s) \label{eqn:reg1}.
\end{equation}
By using the explicit expressions for the coefficients $p_i$ and $q_i$,
one can show that Eq. (\ref{eqn:reg1}) 
is necessary and sufficient to ensure regularity also of Eq. (\ref{eqn:ueqnFull}), i.e. 
Eq. (\ref{eqn:reg1}) is equivalent to 
\begin{equation}
        q_1(r_s)\Omega'(r_s) + q_2(r_s) \Lambda'(r_s) = -\frac{q_3(r_s)}{U(r_s)} \Lambda(r_s). \label{eqn:reg2} 
\end{equation}

Furthermore, the homogeneity of Eq.~(\ref{eqns:new}) in $\Omega$ and $\Lambda$ means that an entire family of solutions can be obtained by rescaling a single solution. In other words, if initial data $\{\Omega'(r_s),\Lambda(r_s),\Lambda'(r_s)\}$ specifies a solution that is well behaved at the spin-1 horizon, then so does $\{J\Omega'(r_s),J\Lambda(r_s),J\Lambda'(r_s)\}$ for any constant $J$. ($J = 0$ gives the spherically symmetric solution.) We will exploit this fact to set $\Omega'(r_s) =1$, so that now spin-1 regularity uniquely constrains $\Lambda'(r_s)$ given $\Lambda(r_s)$ (or vice-versa).

Next we need to discuss the asymptotic behavior of the slowly rotating solutions near spatial infinity. The generic asymptotic solutions are in general linear superpositions of three modes:
\begin{align}
	\Omega' &= \sigma_1 \Omega'_1 + \sigma_2 \Omega'_2  +\sigma_3 \Omega'_3 \label{eq:oasymptotics} \,,\\ 
	\Lambda &= \sigma_1 u_1 + \sigma_2 u_2 + \sigma_3 u_3\,. 
	\label{eq:uasymptotics}
\end{align}
For generic couplings, the mode functions $\{\Omega_i,u_i\}$ behave asymptotically like
\begin{subequations}
	\label{eqn:mode1gen}
\begin{align}
	\Omega'_1 &= -\frac{3}{r^4}-\frac{6c_a c_\sigma M}{(c_\sigma + c_\omega -c_\sigma c_\omega)r^5} + \mathcal{O}\left(\frac{1}{r^6}\right) \,,\\
	u_1 &= \frac{3c_a(1-c_\sigma)}{8(c_\sigma + c_\omega -c_\sigma c_\omega)r^2} + \mathcal{O}\left(\frac{1}{r^3}\right)\,,
\end{align}
\end{subequations}
\begin{subequations}
	\label{eqn:mode2gen}
\begin{align}
	\Omega'_2 &= -\frac{2 c_a (3c_\sigma + c_\omega)}{(c_\sigma+c_\omega - c_\sigma c_\omega)r^5} + \mathcal{O}\left(\frac{1}{r^6}\right)\,, \\
	u_2 &= \frac{1}{r} + \frac{(c_a(2-3c_\sigma) + 2(c_\sigma + c_\omega -c_\sigma c_\omega))M}{2(c_\sigma + c_\omega -c_\sigma c_\omega)r^2} + \mathcal{O}\left(\frac{1}{r^3}\right)\,,
\end{align}
\end{subequations}
\begin{subequations}
	\label{eqn:mode3gen}
\begin{align}
	\Omega'_3 &= \frac{4c_a c_\omega M}{(c_\sigma + c_\omega -c_\sigma c_\omega)r^2} + \mathcal{O}\left(\frac{1}{r^3}\right) \,,\\
	u_3 &= r^2 - \frac{(c_\sigma + c_\omega -c_\sigma c_\omega -2c_a)r}{2(c_\sigma + c_\omega -c_\sigma c_\omega)}+\mathcal{O}\left(r^0\right)\,, \label{eqn:udiverge}
\end{align}
\end{subequations}
where $M$ is the total gravitational mass of the spherical solution. By replacing Eq.~(\ref{eqn:mode3gen}) 
into Eq.~\eqref{metric}, it is clear that the contribution of $\Omega'_3$ to $\Omega$ does not have the right scaling with $r$ 
to be compatible with asymptotic flatness. Similarly, Eq. (\ref{eqn:udiverge}) shows that the aether's $u_\phi$ component diverges as $r^2$ asymptotically. 
Here we wish to impose asymptotic flatness of the metric and asymptotic alignment between the aether and the Killing vector associated with time translations\footnote{Note that 
the latter condition has already been used in the derivation of Eqs.~(\ref{eeq1})-(\ref{eeq2}) in Ref.~\cite{Barausse:2013nwa}.},
which can only be achieved by choosing $\sigma_3=0$.

The preceding analysis can be used to obtain a precise count of the number of free parameters in our solutions. Equations~(\ref{eq:oasymptotics}) and \eqref{eq:uasymptotics} suggest that there are three independent charges, $\sigma_1$, $\sigma_2$, and $\sigma_3$ (in addition to the mass of the spherically symmetric ``seed'' black hole).  Asymptotics fixes $\sigma_3=0$. Regularity of the spin-1 horizon effectively imposes one condition on $\sigma_1$ and $\sigma_2$, which one could always interpret as fixing $\sigma_2$ in terms of $\sigma_1$. $\sigma_1$ is the spin of the slowly rotating black hole, which is therefore the only free parameter apart from the mass of the ``seed'' spherical solution. 

A subtle point in the discussion above and in the counting of the free parameters of the solutions is our implicit assumption that there is only one spin-1 horizon. However, this is not always true. As can be seen from Eq.~(\ref{eqn:sing2}), the roots of the equation $S=0$ depend in a rather complex way on the spacetime structure, the aether configuration, and the value of $s_1$. For appropriate choices of the parameters of the theory, solutions with multiple spin-1 horizons exist. In fact, it is rather easy to find such solutions, even for cases where all of the $c_i$ are rather small and do not have particularly special values. We have empirically discovered that two spin-1 horizons tend to appear when the spin-1 speed is significantly larger that the spin-0 speed. On the other hand,  we conducted a rather thorough search within the experimentally viable parameter space of the theory, and we have not encountered any cases where $S=0$ admits multiple real and positive roots. Therefore, for experimentally viable values of the coupling constants $c_i$, slowly rotating black-hole solutions will indeed have only one spin-1 horizon. Hence, we will not pay particular attention to the possibility of having 2 spin-1 horizons in most of our analysis of the solutions. Nevertheless, we will discuss this issue in more depth in Sec.~\ref{sec:mattingly_solns1}, where we will generate the slowly-rotating counterparts to the explicit spherically symmetric solutions found in Ref.~\cite{Berglund:2012bu} for the special choice $c_\theta = -2c_\sigma$. It turns out that this seed solution does indeed have two spin-1 horizons, and is thus a good example for understanding this feature. 

In general, when more than one spin-1 horizon exists, one expects them to be singular. Recall from our discussion above that in order to render the spin-1 horizon regular we need to impose a local regularity condition. So, if there are multiple horizons, one has to impose multiple local conditions. However, with one regularity condition alone, the solutions are already described by two parameters, the mass and the angular momentum, \footnote{Note that neither of these quantities can be tuned to impose additional
regularity conditions. Indeed, the mass can be set to 1 by rescaling the radial coordinate, while the angular momentum can be set to 1 because it drops out of the field equations
in the slow rotation limit.} leaving no more parameters to tune for imposing further regularity conditions. It is conceivable that all spin-1 horizons may end up being regular once the regularity condition is imposed on the outermost one. This would appear accidental, but could eventually be attributed to some subtle underlying physics. We have considered this possibility and ruled it out. We present the discussion in Appendix \ref{app:manyspin1} (which we recommend reading after Sec.~\ref{solutions}). Hence, we conclude that solutions with more than one spin-1 horizon will exhibit finite area singularities. Note that this is perfectly acceptable from a phenomenological view-point: the outermost spin-1 horizon can be rendered regular with the usual regularity condition discussed above, and hence these finite area singularities will not be ``naked''.

\section{\ae-theory, Ho\v rava gravity and universal horizons}
\label{sec:uhs}

As mentioned in the introduction, Ho\v rava gravity is a theory with a preferred foliation and higher-order dispersion relations. The existence of a preferred foliation allows one to consistently include in the action terms with only two time derivatives but higher-order spatial derivatives, and this is what gives rise to the modified dispersion relations. The presence of these higher-order spatial derivatives modifies the propagators in the ultraviolet end of the spectrum and serves to make the theory power-counting renormalizable \cite{Horava:2009uw,Barvinsky:2015kil}. The presence of a preferred foliation implies that the defining symmetry of the theory is the subset of diffeomorphisms that leave this foliation intact. We only consider here the most general, non-projectable version of the theory, as laid out in Ref.~\cite{Blas:2009qj,Blas:2010hb}, and we do not impose any restriction on the field content or the action other than that imposed by foliation-preserving diffeomorphisms.
Here we are actually only interested in the infrared limit of Ho\v rava gravity and its relation to \ae-theory. Hence, we refrain from giving more details on the general theory and we refer the reader to reviews such as Refs.~\cite{Sotiriou:2010wn,Blas:2014aca,Berti:2015itd}. In fact, in the rest of this paper we will often refer to the infrared limit simply as Ho\^{r}ava gravity; we appeal to brevity to justify this abuse in terminology.

As shown in Ref.~\cite{Jacobson:2013xta},  Ho\v rava gravity can be re-written in a diffeomorphism invariant manner in terms of an \ae ther field that satisfies the following restriction
\begin{equation}
\label{ho}
u_\mu \equiv \frac{\partial_\mu T}{\sqrt{g^{\mu\nu} \partial_\mu T \partial_\nu T}}\,,
\end{equation}
where $T$ is a scalar field whose gradient is always timelike. The action of the infrared part of the theory then becomes formally the same as the action of \ae-theory (\ref{aeaction}). Note, however, that the two theories are not equivalent, as the condition in Eq.~(\ref{ho}) is imposed before the variation. By choosing $T$ as a time coordinate one recovers the preferred foliation and loses part of the diffeomorphism invariance. Foliation-preserving diffeomorphisms become the residual gauge freedom. In the covariant picture, the preferred foliation can be thought of as arising at the level of the solutions, i.e. the level surfaces of $T$ define the preferred foliation. It is worth emphasizing that the field equations become second-order partial differential equations only in the preferred foliation, and are of higher order in other foliations \cite{Jacobson:2013xta,Blas:2009qj}. 

As discussed previously, spherically symmetric solutions of \ae-theory are solutions of Ho\v rava gravity as well, because spherical symmetry makes the \ae ther hypersurface-orthogonal. The converse is not trivially true, but has been shown to hold under the additional assumption of staticity and provided that asymptotically the metric becomes flat and the aether aligns with the timelike Killing vector \cite{Barausse:2013nwa,Barausse:2012ny,Blas:2010hb}. In general, since the condition given by Eq.~(\ref{ho}) is imposed before the variation in Ho\v rava gravity, that theory can admit extra solutions. Also, once spherical symmetry is relaxed, there is no reason why the \ae ther should be hypersurface-orthogonal in \ae-theory, so the solutions of the two theories do not have to match. Indeed, it has been shown in Refs.~\cite{Barausse:2012ny,Barausse:2012qh,Barausse:2013nwa} that \ae-theory does not admit any slowly rotating solutions in which the \ae ther is globally hypersurface-orthogonal.

One can arrive at the same conclusion straightforwardly starting from Eq.~(\ref{eqns:new}).
First, let us demonstrate that slowly rotating Ho\v rava gravity solutions must have $\Lambda=0$ everywhere. Frobenius' theorem ensures that the \ae ther is (globally) hypersurface-orthogonal if and only if 
the twist vector
\begin{equation}\label{eq:twist}
	\omega^\alpha = \frac{1}{\sqrt{-g}}e^{\alpha\beta\mu\nu}u_\beta \nabla_\mu u_\nu=\frac{1}{\sqrt{-g}}e^{\alpha\beta\mu\nu}u_\beta \partial_\mu u_\nu
\end{equation}
($e^{\alpha\beta\mu\nu}$ being the antisymmetric symbol) vanishes everywhere. 
Indeed, Frobenius' theorem states that the vanishing of the three-form $u_{[\alpha} \nabla_\beta u_{\gamma]}$ is a necessary and sufficient condition for $u_\alpha$ to be hypersurface-orthogonal. The twist vector is just the (Hodge) dual of this three-form, and is related to the vorticity tensor, $\omega_{\alpha\beta}$, defined in Eq. (\ref{eqn:vort}) by
\begin{equation}
	\omega_{\alpha\beta} = -\frac{1}{2}\sqrt{-g}e_{\alpha\beta\mu\nu}u^\mu \omega^\nu.
\end{equation}
With our  ans\"atze for the metric and \ae ther, the nonvanishing components of the twist are 
\begin{align}
	&\begin{aligned}
	\omega^t = \epsilon\left(\frac{1-A^2f}{r^2Af}\right) \Lambda \cos\theta + \mathcal{O}(\epsilon^3) \label{eq:twist1}
	\end{aligned}\\
	&\begin{aligned}
	\omega^r = -\epsilon\left(\frac{U}{r^2AB}\right) \Lambda \cos\theta + \mathcal{O}(\epsilon^3) \label{eq:twist2}\\
	\end{aligned}\\
	&\begin{aligned}
	\omega^\theta =\,\, &\epsilon\left[\frac{((1-A^2f)A'-A^3f')\Lambda + UA\Lambda'}{2r^2A^2B}\right] \sin\theta \\ &+ \mathcal{O}(\epsilon^3)\,. \label{eq:twist3}
	\end{aligned}
\end{align}
where $U$ is as defined in Eq. (\ref{eqn:sing1}). As can be seen, the twist vanishes globally if and only if  $\Lambda=0$ everywhere.

Let us now integrate Eq. (\ref{eqn:oeqnFull}) to give
\begin{align}
	\Omega'(r) = \frac{1}{\mathcal{Q}(r)}\left(\kappa + \int^r \mathcal{Q}(\rho)\mathcal{J}(\rho)d\rho\right),
	\label{eqn:Omegasol}
\end{align}
where $\kappa$ is an integration constant,
\begin{equation}
	\mathcal{Q}(r) = \exp\left[\int^r \frac{p_1(\rho)}{S(\rho)} d\rho\right]
	\label{eqn:integconst}
\end{equation}
and 
\begin{equation}
\mathcal{J}(r) = \frac{1}{S(r)}\left(p_2(r)\Lambda'(r)+ \frac{p_3(r)}{U(r)}\Lambda(r)\right).
	\label{eqn:Omegasource}
\end{equation}
Inserting this back into (\ref{eqn:ueqnFull}) gives an inhomogenenous, linear, second-order, integro-differential equation for $\Lambda$:
\begin{align}
		\Lambda'' = \frac{1}{S}\Big[q_2 \Lambda' &+ \frac{q_3}{U}\Lambda +  \frac{q_1}{\mathcal{Q}}\int^r \frac{\mathcal{Q}}{S}\left(p_2 \Lambda'(\rho) + \frac{p_3}{U}\Lambda(\rho) d\rho\right)\Big] \nonumber \\ &+ \frac{\kappa q_1}{\mathcal{Q}S}. \label{eqn:integro}
\end{align}

The Ho\v rava-gravity  solution  $\Lambda(r) = 0$ is obtained only when the inhomogeneous term, $\kappa q_1/(S\mathcal{Q})$, is identically zero. For this to happen  $q_1/S$ has to be zero, as $\mathcal{Q}$ cannot be made to diverge for any value of the coupling constants. $q_1/S$  vanishes as $c_\omega \rightarrow \infty$ or $c_a \rightarrow 0$.  The first possibility is particularly interesting and we will discuss it in the next subsection. The second case, $c_a =0$, is special as both the spin-0 and the spin-1 mode have diverging speeds at this limit. Moreover, static, spherically symmetric solutions are known in closed form in \ae-theory for this choice of $c_a$~\cite{Berglund:2012bu}, and we will return to it in Sec. \ref{sec:cazero}. 

Recall that, by definition, a universal horizon is a compact hypersurface that encloses the central singularity and to which the \ae ther is orthogonal. 
On this surface all components of the twist vector [Eq.~\eqref{eq:twist}] would have to vanish. Based on our previous analysis and the result of Refs.~\cite{Barausse:2012ny,Barausse:2012qh,Barausse:2013nwa}, which prove that the twist cannot vanish globally, it may be tempting to conclude that slowly rotating solutions in \ae-theory cannot possess universal horizons. However, to prove this beyond doubt, one needs to actually show that the  \ae ther does not become orthogonal to \emph{any} hypersurface, without necessarily being globally hypersurface-orthogonal. 
In our setting  one has to show that the vorticity cannot vanish even on a single hypersurface of constant $r$. 
This is one of the main motivations for finding slowly rotating solutions.

Indeed, once slowly rotating solutions are available, checking whether universal horizons exist is straightforward. 
In more detail, Eq.~\eqref{eq:twist1} shows that 
in order for $r=r_u$ to be a universal horizon, one must have $\Lambda(r_u) = 0$. This is because the combination $1-A^2 f\propto u_r$
never vanishes (for generic viable values of the coupling constants $c_i$) in spherically symmetric and static black holes~\cite{Barausse-etal-PRD:2011}. By looking at 
Eqs.~\eqref{eq:twist2}--\eqref{eq:twist3}
it is then clear that for the other components of the 
twist to vanish at $r=r_u$ we must have either $U(r_u)=0$ -- which happens if and only if $r=r_u$ is the universal horizon of the 
spherically symmetric and static solution -- or $\Lambda'(r_u) = 0$. 

\subsection{The $c_\omega\to\infty$ limit}
\label{sec:limit}

Reference~\cite{Jacobson:2013xta} argued that solutions to \ae-theory converge to Ho\v rava gravity solutions in the $c_\omega\to\infty$-limit (provided that they remain regular in that limit). 
In this section we will briefly review the main argument of Ref.~\cite{Jacobson:2013xta}, and check its validity in the case of slowly rotating solutions.

The action of \ae-theory, Eq.~(\ref{aeaction}), contains the vorticity-dependent term $\sqrt{-g} c_\omega \omega_{\mu\nu}\omega^{\mu\nu}$. Variation of this term with respect to $g^{\mu\nu}$ (keeping $u_\mu$ fixed)
yields the same stress energy tensor as the electromagnetic field, i.e. the Einstein equations become
\begin{align}\label{Tw}
\delta S=&\delta \left(\int c_\omega \omega_{\mu\nu}\omega^{\mu\nu} \sqrt{-g}  d^4 x\right)
+\mbox{terms independent of $c_\omega$}\notag\\&
= 2 c_\omega \left(\omega_{\mu\alpha} \omega_{\nu}^{\phantom{a}\alpha}-
\frac14 g_{\mu\nu}\omega_{\alpha\beta} \omega^{\alpha\beta}\right)\sqrt{-g}\delta g^{\mu\nu}\notag\\&+\mbox{terms independent of $c_\omega$.}
\end{align}
If this contribution to the (generalized) Einstein's equations is to remain finite in the limit $c_\omega\to \infty$, then
\begin{equation}
\omega_{\mu\alpha} \omega_{\nu}^{\phantom{a}\alpha}-\frac14 g_{\mu\nu}\omega_{\alpha\beta} \omega^{\alpha\beta}\to0.
\end{equation}
Contraction of this equation with $u^\mu$ shows that $\omega_{\mu\nu}\to 0$, because $\omega_{\mu\nu} u^\nu=0$, and $\omega_{\mu\nu} \omega^{\mu\nu}>0$ 
unless $\omega_{\mu\nu}=0$. (Both of these expressions are obvious if one notes that the vorticity
definition, Eq.~\eqref{eqn:vort}, can be re-written as $\omega_{\mu\nu}=h^{\alpha}_{\mu}h^\beta_{\nu} \nabla_{[\beta} u_{\alpha]}$.)
Therefore, the vorticity-free solutions are the only regular ones in this limit. Reference~\cite{Jacobson:2013xta} then argues, by a simple example, that the \ae ther's field equations do not impose any additional restrictions in the  $c_\omega\to \infty$ limit, and that the equations and the solutions should consequently converge to those of Ho\v rava gravity
in that limit.

A subtlety that has been missed in Ref.~\cite{Jacobson:2013xta} comes from the fact that if $\omega_{\mu\nu}\sim c_\omega^{-1/2}$, then it can still vanish in the $c_\omega\to \infty$ limit and yet give a finite contribution to the Einstein equations. Indeed, 
the $c_\omega$ dependent terms in Eq.~\eqref{Tw} are exactly the difference between the Einstein equations in
\ae-theory and Ho\v rava gravity~\cite{Barausse:2012qh,Barausse:2013nwa}. Therefore, these terms should vanish in the limit $c_\omega\to\infty$ if Ho\v rava-gravity solutions are to be recovered from \ae-theory ones. One cannot assess if $\omega_{\mu\nu}$ vanishes faster than $c_\omega^{-1/2}$ or not without considering explicitly the \ae ther's equations in this limit. Hence, one cannot actually argue without doubt that \ae-theory solutions will converge to Ho\v rava gravity solutions on the basis of Eq.~\eqref{Tw} alone.

By varying instead the \ae-theory action with respect to $u_\mu$ and enforcing the unit norm constraint $u^2=1$, one obtains the following contribution to the \ae ther equations
from the $\sqrt{-g} c_\omega \omega_{\mu\nu} \omega^{\mu\nu}$ term: 
\begin{align}\label{ae}
\delta S&=\delta \left(\int c_\omega \omega_{\mu\nu} \omega^{\mu\nu} \sqrt{-g}  d^4 x\right)
+\mbox{terms independent of $c_\omega$}\notag \\&=
2 c_\omega h^{\mu}_{\alpha}\left(\nabla_\nu \omega^{\alpha\nu}-\omega^{\alpha\nu}a_\nu\right)\sqrt{-g}\delta u_\mu\notag\\&+\mbox{terms independent of $c_\omega$.}
\end{align}
Now, in the $c_\omega\to \infty$ limit, regularity of this contribution yields a differential equation for $\omega_{\mu\nu}$. Provided that a suitable combination of asymptotic, boundary and/or initial conditions are imposed, one can use this equation to argue that the vorticity should be $\omega_{\mu\nu}={\cal O}(1/c_\omega)$, which would indeed be enough to ensure that the $c_\omega$-dependent terms in Eq.~\eqref{Tw} disappear as $c_\omega\to\infty$. This highlights the importance of both the actual structure of the full set of field equations and the appropriate choice of asymptotic, boundary and/or initial conditions to obtain the desired result. For the sake of clarity, in Appendix \ref{app:cw} we present an elementary example that shares most of the structure of Eqs.~\eqref{Tw} and \eqref{ae}, and yet fails to have the desired limit precisely because it allows for what would be the analog of solutions with $\omega_{\mu\nu}\sim c_\omega^{-1/2}$ scaling.

We can now focus on slowly rotating solutions and attempt to apply the rationale above to argue that \ae-theory solutions converge to the Ho\v rava gravity one as $c_\omega\to \infty$. However, there is a complication: since the vorticity vanishes for the spherically symmetric ``seed'' solutions, the term in Eq.~\eqref{Tw} vanishes to first order in rotation for any value of $c_\omega$.\footnote{This should also act as a note of caution that the slow rotation approximation might introduce spurious solutions.} Hence, the \ae ther equation is the only equation that determines the vorticity and thus its behavior in the $c_\omega \to \infty$ limit. Indeed, 
in this limit one has
$p_1/S \rightarrow -{4}/{r} + {B'}/{B}$ in Eq.~(\ref{eqn:integconst}), which turns Eq.~(\ref{eqn:Omegasol}) into
\begin{equation}
	\Omega'(r) = \frac{B(r)}{r^4} \left[\kappa + \int^r \frac{\mathcal{J}(\rho)\rho^4}{B(\rho)} d\rho\right],
	\label{eqn:Omegalimit}
\end{equation}
where $\mathcal{J}(\rho)$ is defined in Eq.~(\ref{eqn:Omegasource}). 
Also, $q_1/S \rightarrow 0$ in Eq.~(\ref{eqn:integro}),  while $q_2/S$ and $q_3/(US)$ converge to finite expressions. 
Therefore, the \ae ther potential $\Lambda$ fully decouples from the frame-dragging potential $\Omega'$, 
and Eq.~(\ref{eqn:integro}) becomes a homogeneous, second-order differential equation. As a result, $\Lambda$ is not necessarily trivial, at least not without additional input such as boundary conditions, and the corresponding frame-dragging in Eq.~(\ref{eqn:Omegalimit}) is not necessarily that of the slowly rotating Ho\v{r}ava solution, $\Omega'(r) = \kappa B(r)/r^4$, found in Ref.~\cite{Barausse:2012qh}. 

Perhaps it is more illuminating to go back to Eqs.~\eqref{eeq1} and \eqref{eeq2}. Combining them linearly so
as to eliminate $\psi'$, one obtains an equation that, in the limit  $c_\omega\to\infty$, does not depend on $\psi''$ either. More precisely,
 one obtains
\begin{align}
&\lambda+\lambda' L_1+\lambda'' L_2={\cal O}\left(\frac{1}{c_\omega}\right)\,,\label{masterEq}\\
&L_1=\frac{r^2 \left(A^2 f+1\right)}{{8 A^3 B^3}}\times\notag \\&\;\times  \left\{A \left(A^2 f+1\right) B'-4 B   \left[\left(A^2 f-1\right) A'+A^3 f'\right]\right\}\\
&L_2=-\frac{r^2\left( A^2 f+1\right)^2}{8 A^2 B^2}\,.
\end{align}
This is precisely the equations one would obtain by looking at Eq.~(\ref{ae}) as $c_\omega \to \infty$. 

Reference~\cite{Jacobson:2013xta} considered the slowly rotating case explicitly as an example, starting from Eqs.~\eqref{eeq1} and \eqref{eeq2}, and argued  that \ae-theory slowly rotating solutions converge to Ho\v rava gravity ones as $c_\omega\to \infty$, without the need to impose any condition other than asymptotic flatness. This is in direct contradiction to the result of our analysis above. According to Ref.~\cite{Jacobson:2013xta}, $d_3$ in Eq.~\eqref{eeq1} scales as $c_\omega^2$, whereas $d_1, d_2$ and $d_4$ only scale as $c_\omega$. If this is the case, as $c_\omega$ is taken to infinity, regular solutions of Eq.~(\ref{eeq1}) will have to satisfy $\lambda'(r)$ = 0. Together with asymptotic flatness, this means that $\lambda(r) =\Lambda(r)=0$. This reasoning thus leads to the known slowly rotating Ho\v{r}ava solution of Ref.~\cite{Barausse:2012qh}. 

Clearly, the crux of this argument rests on $d_3$ growing faster than $\{d_1,d_2,d_4\}$ as $c_\omega \rightarrow \infty$.
However, from the explicit expressions given in Appendix \ref{app:coeffs}, it follows that all of the $d_i$ are in fact linear in $c_\omega$, whereas all of the $b_i$ coefficients are independent of $c_\omega$. This is in complete agreement with our analysis above and Eqs.~(\ref{eqn:Omegalimit}) and (\ref{masterEq}).

Having established that it is Eq.~(\ref{masterEq}) that determines whether \ae-theory solutions converge to Ho\v rava ones in  the $c_\omega\to \infty$ limit, we can now return to it and solve it at lowest order in $1/c_\omega$, where the right-hand side
is exactly zero.
Because asymptotic flatness was implicitly used to derive Eqs.~\eqref{eeq1} and \eqref{eeq2}, we need to
 impose $\lambda\to0$ as $r\to\infty$. Nevertheless, even with this boundary condition, 
 the Ho\v rava gravity solution $\lambda=0$ is not the only solution to Eq.~\eqref{masterEq} if no other boundary condition is added.
For slowly rotating stars, 
in spite of the spacetime being non-vacuum, 
Eq.~\eqref{masterEq} still holds\footnote{Note however that for stars one has $f(r) A(r)^2=1$~\cite{Eling:2006df,Eling:2007xh}.},  because 
it comes from the  $c_\omega\to\infty$ limit of the
\ae ther equations alone.
Regularity at the center imposes $\lambda'=0$ at $r=0$.\footnote{To see this,
one needs to transform to Cartesian coordinates (since spherical coordinates are singular at the center),
and note that $\lambda={\cal O}(r)={\cal O}(\sqrt{x^2+y^2+z^2})$ is not differentiable at $r=x=y=z=0$.} 
Moreover, it follows from Eq.~\eqref{masterEq} that regularity of
$\lambda$ at $r=0$ also requires $\lambda(r=0)=0$. (To see this, one
can simply replace $r=0$ in Eq.~\eqref{masterEq}, while assuming
a regular $\lambda$.) These two conditions are already enough to
select $\lambda=0$ as the unique solution, even without 
using the asymptotic-flatness boundary condition.
Then, taking into account how finite-$c_\omega$ corrections enter Eq.~\eqref{masterEq}, it follows that $\lambda(r)={\cal O}(1/c_\omega)$. Finally, replacing this solution in either Eq.~\eqref{eeq1} or Eq.~\eqref{eeq2}, one has
\begin{equation}\label{combo}
r B' \Omega'-B \left(r \Omega''+4 \Omega '\right)={\cal O}\left(\frac{1}{c_\omega}\right)\,,
\end{equation}
from which one obtains
\begin{equation}\label{rot_soln}
\Omega(r)=- 12 J\int_{}^r \frac{B(\rho)}{\rho^4} d\rho +\Omega_0+{\cal O}\left(\frac{1}{c_\omega}\right)\,,
\end{equation}
with $J$ the solution's spin and $\Omega_0$ an integration constant that can be set to zero without loss of generality (as it can be
made to vanish with a coordinate change $\phi\to \phi+\Omega_0 t$). This is indeed the Ho\v rava gravity slowly rotating solution up to remainders ${\cal O}\left({1}/{c_\omega}\right)$~\cite{Barausse:2012qh,Barausse:2013nwa}.

Let us now turn our attention to black-hole solutions, where no regularity condition at the center can be imposed. This condition is actually replaced by
the requirement that the spin-1 horizon be regular, as we will now demonstrate. Recall that the spherical ``seed'' solution possesses a universal horizon where $1+f A^2=0$.
In the limit $c_\omega\to \infty$, the universal horizon actually coincides with the spin-1 horizon because the spin-1 mode travels at infinite speed [cf.~Eq.~(\ref{eqn:s1mode})].
Let us solve  the $c_\omega\to\infty$-limit of Eq.~\eqref{masterEq} perturbatively near the spin-1/universal horizon, whose radius we denote by $r_u$.
To this end, we expand $f(r)=f_0+f_1 (r-r_u)+{\cal O}(r-r_u)^2$, $B(r)=B_0+B_1 (r-r_u)+{\cal O}(r-r_u)^2$, and
$A(r)=A_0+A_1 (r-r_u)+{\cal O}(r-r_u)^2$. Since $1+f A^2=0$ at $r=r_u$, one has $f_0=-1/A_0^2$.
With these expansions, the coefficients $L_1$ and $L_2$ become
\begin{gather}
L_1=-\frac{\left(r-r_u\right) r_u^2 \left(A_0^3 f_1-2 A_1\right){}^2}{2 A_0^4 B_0^2}+{\cal O}(r-r_u)^2\\
L_2=-\frac{\left(r-r_u\right){}^2 r_u^2 \left(A_0^3 f_1-2 A_1\right){}^2}{8 A_0^4 B_0^2}+{\cal O}(r-r_u)^3
\end{gather}
and the general solution to the $c_\omega\to\infty$-limit of Eq.~\eqref{masterEq} is
\begin{multline}\label{sol}
\lambda=\left(r-r_u\right){}^{\chi-\frac{3}{2}} \left\{k_2 [1+{\cal O}(r-r_u)] \left(r-r_u\right){}^{-2 \chi}\right.\\\left.+k_1 [1+{\cal O}(r-r_u)]\right\}
\end{multline}
where $k_1$ and $k_2$ are integration constants, and 
\begin{equation}
\chi=\frac{A_0^2 B_0 \sqrt{\frac{9 r_u^2 \left(A_0^3 f_1-2 A_1\right){}^2}{A_0^4 B_0^2}+32}}{4 A_1 r_u-2
   A_0^3 f_1 r_u}\,.
\end{equation}
One can easily verify that $|\chi|>3/2$. Therefore,
if $\chi<0$, in order for $\lambda$ [or $u_\phi= (1+f A^2) \lambda \sin^2\theta/(2A) \approx 
  \left(r-r_u\right) \left(A_0^3 f_1-2 A_1\right) \sin ^2\theta  \lambda\left(r_u\right) /(2 A_0^2)$]
to be finite at $r=r_u$, we must have $k_1=0$. Likewise, if $\chi>0$ we must have $k_2=0$ to ensure $\lambda$ (and $u_\phi$) are finite. In either
case, the finite branch of the solution given by Eq.~\eqref{sol} vanishes at $r=r_u$, i.e. regularity requires $\lambda=0$
at the universal/spin-1 horizon.
If $|\chi|<5/2$, however, $\lambda=0$ at $r=r_u$ does not ensure that $\lambda'$ is finite there. In fact, by using the solution
given by Eq.~\eqref{sol} and reasoning like 
we just have for $\lambda$, it is easy to see that $\lambda'$ is finite at $r=r_u$ if and only if it is zero there.
Therefore, for $|\chi|<5/2$, regularity imposes $\lambda=\lambda'=0$ at $r=r_u$ and thus selects the unique trivial solution $\lambda=0$.
If instead $|\chi|\geq 5/2$, regularity only imposes $\lambda=0$ at $r=r_u$, but together with asymptotic flatness
($\lambda\to0$ as $r\to\infty$) this still selects the unique solution $\lambda=0$.\footnote{Note that
we can only conclude that the solution to the boundary value problem given by Eq.~\eqref{masterEq} and $\lambda(r_u)=\lambda(\infty)=0$
is unique because $L_2<0$ everywhere for $r>r_u$. Indeed, this ensures that if there is a local extreme value for
a solution $\lambda$, then it will be a local minimum (maximum) if $\lambda>0$ ($\lambda<0$). This is enough to conclude that
there cannot be any non-trivial solution satisfying  $\lambda(r_u)=\lambda(\infty)=0$.}
We can therefore conclude that irrespective of the value of $\chi$, the 
only regular asymptotically flat solution to the $c_\omega\to\infty$-limit of  Eq.~\eqref{masterEq} is
$\lambda=0$. As in the case of
slowly rotating stars, one can then restore the remainders ${\cal O}\left({1}/{c_\omega}\right)$ on the right-hand side of 
Eq.~\eqref{masterEq} and conclude that $\lambda={\cal O}\left({1}/{c_\omega}\right)$. Finally, by replacing in either 
Eq.~\eqref{eeq1} or Eq.~\eqref{eeq2}, one  obtains again Eq.~\eqref{rot_soln}, which matches the Ho\v rava gravity solution for $c_\omega\to \infty$.

In conclusion, \ae-theory slowly rotating solutions that describe  black holes and stars do indeed converge to Ho\v rava-gravity solutions 
for $c_\omega\to\infty$, provided that suitable regularity conditions are imposed.  It is, however, conceivable that more generic slowly rotating solutions (e.g. around wormhole solutions) might not exhibit the same behavior, as in the absence of additional boundary conditions, Eq.~\eqref{masterEq} admits solutions different from the Ho\v rava gravity solution $\lambda(r)=0$
(even with the asymptotic-flatness condition $\lambda(\infty)=0$).

\section{Slowly Rotating Solutions}
\label{solutions}

\subsection{Solutions in the small-coupling limit}
\label{sec:small_couplings}

The dimensionless coupling constants of \ae-theory are constrained to be
$|c_\theta|,|c_a|,|c_\sigma|,|c_\omega|\lesssim \mbox{a few }\times 0.01 $ by gravitational observations 
(especially binary pulsars~\cite{Yagi:2013qpa,Yagi:2013ava}). In this small-coupling regime, the propagation speeds of the spin-0, spin-1 and spin-2 graviton polarizations become
\begin{align}
	s^2_0 &= \frac{c_\theta + 2c_\sigma}{3c_a} + {\cal O}(c) \\
	s^2_1 &= \frac{c_\sigma+c_\omega}{2c_a} + {\cal O}(c),\\
	s^2_2 &= 1 + {\cal O}(c)\,,
\end{align}
where ${\cal O}(c)\equiv {\cal O}(c_\theta,c_a,c_\sigma,c_\omega)$, while the spherically symmetric black-hole solutions
reduce to the Schwarzschild spacetime plus corrections, $B = 1 + O(c)$ and $f=1-r_0/r +O(c)$, 
where $r_0=2 M$ ($M$ being the black-hole mass). The \ae ther potential, $A(r)$, obeys 
the small-coupling equation
\begin{equation}
	A''(r) = \frac{P(r)}{Q(r)}+{\cal O}(c),  \label{eqn:Asmallcoupling}
\end{equation}
where
\begin{align}
	P(r) = \,\,&2\left(\frac{r}{r_0}\right)^4\bigg(-2+(1+s_0^2)U\bigg)A'{}^2 \nonumber\\ &- 2\left(\frac{r}{r_0}\right)^3 \bigg((s_0^2-1) + (s_0^2+1)(1+f)A^2 \nonumber\\ &+ (s_0^2-1)f A^4\bigg)AA' - 2 s_0^2 \left(\frac{r}{r_0}\right)^2 (2-U)UA^2
\end{align}
and
\begin{align}
	Q(r) =\,\,&A \left(\frac{r}{r_0}\right)^4\bigg((s_0^2-1)U^2 + 4fA^2\bigg) \nonumber\\ &\propto g_{tt} + (s_0^2-1)u_t^2, \nonumber
\end{align}
where the prime denotes differentiation with respect to $r$.
Note that the proportionality in the last equation shows that $Q$ vanishes at the spin-0 horizon, which is therefore (in general) a singular point.
To ensure regularity of $A$, one must therefore impose that $P$ also vanishes at the spin-0 horizon, which results in a relation
between the values of $A$ and $A'$ there.
Regular spherically symmetric and static solutions in the small-coupling approximation to \ae-theory and Ho\^{r}ava gravity were first studied in Ref.~\cite{blas-sibiryakov-prd:2011}. Eq. (\ref{eqn:Asmallcoupling}) above is equivalent to Eq. (38) of this reference.

Let us now consider the field equations to first order in rotation. In the small-coupling limit, Eqs. (\ref{eqn:oeqnFull}) and (\ref{eqn:ueqnFull}) reduce to 
\begin{subequations}
\begin{align}
\Omega'' &= -\frac{4}{r}\Omega' + {\cal O}(c) \,\,\,\Longrightarrow\,\,\, \Omega'(r) = \kappa/r^4+{\cal O}(c), \label{eqn:oeqn}\\ 
\Lambda'' &= \frac{1}{S}\left(h_1\Lambda' + \frac{h_2}{U} \Lambda + \kappa h_3\right)+{\cal O}(c), \label{eqn:ueqn}
\end{align}
\label{eqn:oueqns}
\end{subequations}
where $\kappa$ is an integration constant,
\begin{align}
r^2 h_1(r) =\,\, &-4A^2 + 2(1-s_1^2)U\Big(A^2 - r^2 \frac{d 
\log A}{dr} \nonumber\\ &+r^2fAA'\Big)\\
r^2 h_2(r) =\,\, &8 rfA(A^2-1)A' + 4r^2f\Big(2A'{}^2+(U-2)AA''\Big) \nonumber \\&+ 8A^2U- (1-s_1^2)\Bigg\{\frac{2}{r^2A^2}\Big[2 r^4 A'{}^2-\nonumber\\ & r^4  f A^2 A'{}^2  -r^2 A^3 A' +3 r^2 f A^5 A' \nonumber\\ &+(2-r) A^6+r A^4 (r^3 f^2A'{}^2+4 r-1)\Big] \nonumber \\ &+\frac{A''}{A}U(U-2)\Bigg\}U\\
r^2 h_3(r) =\,\,& f'A^3 - (U-2)A', 
\end{align}
where $U$ and $S$ are as defined in Eqs.~\eqref{eqn:sing1} and \eqref{eqn:sing2}. Note that whereas $h_1$ and $h_3$ depend only on $A$ and $A'$, $h_2$ also depends explicitly on $A''$.

To arrive at Eq. (\ref{eqn:ueqn}), we have first solved Eq. (\ref{eqn:oeqn}) for $\Omega'$.
Setting the integration constant $\kappa$ to $12 J$ in this solution, 
we recover the slowly rotating Kerr solution of GR with angular momentum $J$.  
Nevertheless, one can also set  $\kappa$ to 1 without loss of generality, simply by rescaling the variables: $\{\Lambda\to \kappa\Lambda, \Omega' \to \kappa\Omega'\}$; we will always make this choice when solving the field equations numerically. 
In more detail, by replacing the solution for  $\Omega'$ into
Eq. (\ref{eqn:ueqnFull}), we arrive at Eq. (\ref{eqn:ueqn}), which depends
on the coupling constants only through the speed of the spin-1 mode, $s_1^2$, and implicitly on the spin-0 mode speed through $A(r)$. 

Equation (\ref{eqn:ueqn})  appears to have singular points at the spin-1 and universal horizons of the spherically symmetric seeds, where
 $S$ and $U$ vanish respectively. We discussed previously that only $S=0$ is a true singular point, while $U=0$ can be shown to be regular; this is true even beyond the small-coupling regime. In the current small-coupling setting, 
this can be demonstrated explicitly by using Eq.~\eqref{eqn:Asmallcoupling} for the spherically symmetric \ae ther. Indeed, one can add a multiple of  Eq.~\eqref{eqn:Asmallcoupling}  to the right hand side of Eq.~\eqref{eqn:ueqn} in the following way,
\begin{multline}
\Lambda'' = \frac{1}{S}\left(h_1\Lambda' + \frac{h_2}{U} \Lambda + h_3\right)
\\-\left(A'' - \frac{P}{Q}\right) \frac{2 f Q}{r^4U S} \Lambda(r)
+{\cal O}(c) \label{eqn:ueqnother0}\,,
\end{multline}
and the resulting equation (when $\kappa$ is set to 1 as discussed above) reads
\begin{equation}
	\Lambda'' = \frac{1}{S}\left(h_1\Lambda' + \bar{h}_2 \Lambda +h_3\right)\label{eqn:ueqnother}\,,
\end{equation}
where the coefficient $h_1$ and $h_3$ are unchanged, while $\bar{h}_2$ is given by
\begin{align}
-r^4A^2\bar{h}_2 =\,\, &r^4 (s_1^2-1) A A'' - 4 r^4 (s_1^2-1) A'{}^2  \nonumber\\ &- 2 r^4 f(2 s_0^2-s_1^2+3) A^2 A'{}^2 \nonumber\\ &- 2 r A^4
   \Big[((s_1^2-1) r^2 f^2 A'{}^2 -2 r f s_0^2 \nonumber\\ &+(4 r-1) s_1^2+1\Big] +2 r^2 A^3 \Big[(s_0^2+1) r^2 f A'' \nonumber \\ &+(2
   rf s_0^2+2 r+s_1^2-3) A'\Big] \nonumber \\ &+ r^2 f A^5 \Big[(2 s_0^2-s_1^2-1)r^2f A'' \nonumber \\ &+(4 r (s_0^2-1)-6
   s_1^2+6) A'\Big] \nonumber\\ &-2 A^6 \Big[2 s_0^2 r^2f^2 - (r-2) (s_1^2-1)\Big],
\end{align}
which is regular everywhere except for $r=0$. (Note that $A(r)$ cannot vanish at any $r$ because the spherically symmetric aether is required to be future-directed and timelike.) 
The singular point at the spin-1 horizon $r=r_s$ survives these manipulations, and indeed it can only be avoided by imposing the regularity condition
\begin{equation}\label{reg}
h_1\Lambda' + \bar{h}_2 \Lambda +h_3= 0 
\end{equation}
at $r=r_s$. 

\subsubsection*{Numerical implementation; asymptotics}

To solve the field equations numerically, we set $r_0=1$ by rescaling the radial coordinate, and solve Eq.~\eqref{eqn:Asmallcoupling} for $A$, while
imposing regularity at the spin-0 horizon and matching to an asymptotically flat solution~\cite{Barausse-etal-PRD:2011}
\begin{equation}
 A = 1 + \frac{1}{2r} + \frac{a_2}{r^2} + \frac{a_2 - 16}{r^3} + {\cal O}\left(\frac{1}{r^4}\right),
\end{equation}
where $a_2$ is a secondary \ae ther charge (i.e. $a_2$ is fixed once $r_0$ and the coupling constants are fixed).
Numerical solutions that satisfy this asymptotic behavior are found via shooting, just as in Refs.~\cite{blas-sibiryakov-prd:2011,Barausse-etal-PRD:2011}.

We  then insert the numerical solution for $A$ into Eq. (\ref{eqn:ueqnother}) and then integrate the resulting differential equation numerically. (In practice, we interpolate
the solution for $A$ when using it in Eq. (\ref{eqn:ueqnother}), see below for a discussion of how
our results depend on the interpolation scheme.)
We seek solutions for $\Lambda$ that are regular at the spin-1 horizon [c.f. Eq.~\eqref{reg}] and asymptotically flat. The generic asymptotic solution of Eq. (\ref{eqn:ueqnother}) as $r\rightarrow \infty$ is given by 
\begin{equation}
	\Lambda = \sigma_1 \Lambda_1 + \sigma_2 \Lambda_2 + \Lambda_3,
	\label{asympdec}
\end{equation}
where 
\begin{align}
	\Lambda_1 &= r^2 + \left(\frac{1}{2}-\frac{1}{2s_1^2}\right)r + {\cal O}(1), \\
	\Lambda_2 &= \frac{1}{r} + \left(\frac{1}{2}+\frac{1}{4s_1^2}\right)\frac{1}{r^2} + {\cal O}\left(\frac{1}{r^3}\right), \\
	\Lambda_3 &= \frac{1}{16 s_1^2 r^2} + {\cal O}\left(\frac{1}{r^3}\right)\,,
\end{align}
and $\sigma_1$, $\sigma_2$ are integration constants.
Generic solutions will therefore diverge asymptotically, whereas asymptotically flat ones 
are those for which the divergent mode $\Lambda_1$ is absent, i.e. $\sigma_1=0$. In order to select asymptotically flat solutions, we use again a shooting method, using $\Lambda(r_s)$ at the spin-1 horizon as the shooting parameter. For a given choice of $\Lambda(r_s)$, Eq.~\eqref{reg} determines $\Lambda'(r_s)$, thus fixing all the initial data needed to compute $\Lambda(r)$ from $r=r_s$ up to large radii $r \gg r_s$.

In practice, because Eq.~\eqref{eqn:ueqnother} presents a singular point at $r=r_s$, we cannot start the numerical integration exactly from there. Instead we consider the
Taylor expansion
\begin{multline}
	\Lambda(r) = \Lambda(r_s) + \Lambda'(r_s)(r-r_s) + \frac{1}{2}\Lambda''(r_s)(r-r_s)^2  \\ + {\cal O}(r-r_s)^3,
\end{multline}
where $\Lambda''(r_s)$
is determined in terms of the initial data $\Lambda(r_s)$ and $\Lambda'(r_s)$ by solving Eq.~(\ref{eqn:ueqnother}) perturbatively.
This perturbative solution is then used to determine initial data for the numerical integration at
$r=r_s +\delta$ with $\delta = 10^{-8}$.  

In more detail, to find the unique value of $\Lambda(r_s)$ that gives $\sigma_1=0$, we first find two values $\Lambda_1(r_s) < \Lambda_2(r_s)$, such that one gives a solution with $\sigma_1 < 0$, and the other gives $\sigma_1 > 0$. This determines a bracket $(\Lambda_1(r_s),\Lambda_2(r_s))$ that contains the sought-after $\Lambda(r_s)$. Then, just as in standard bisection, we systematically shrink this bracket until we settle on a value $\Lambda(r_s)$ that gives a sufficiently small $\sigma_1$. 
The threshold for $\sigma_1$ is chosen to be\footnote{Note that this is much larger than our machine precision because we use 30 significant digits.}  $|\sigma_1| < 10^{-16}$.
This procedure determines the correct initial data $(\Lambda(r_s),\Lambda'(r_s))$ that yield an asymptotically flat solution. 
These initial data are also used to integrate Eq. (\ref{eqn:ueqnother}) inward from the spin-1 horizon, down to very small distances from the central singularity at $r=0$. 

The value of $\sigma_1$ for a given solution is extracted by fitting to the functional form $\alpha_1 r^2 + \alpha_2 r + \alpha_3 + \alpha_4/r + \alpha_5/r^2$, where clearly $\sigma_1=\alpha_1$, at large radii $r>r_\infty$. (Note that we typically choose $r_\infty=1000$, although our results are robust against this choice.) 
This procedure also allows testing the consistency of our results, because the asymptotic solutions in Eq.~\eqref{asympdec} imply that the extracted coefficients $\alpha_4$ and $\alpha_5$ must satisfy 
\begin{equation}
        \alpha_5 - \frac{1}{16 s_1^2} = \left(\frac{1}{2}+\frac{1}{4s_1^2}\right) \alpha_4, \label{check}
\end{equation}
while $\alpha_2$ and $\alpha_3$ should vanish. We have checked that Eq.~\eqref{check} is satisfied by our numerical solutions to within an accuracy of $10^{-7}$, and that $\alpha_2$ and $\alpha_3$ are zero to within an accuracy of $10^{-16}$ and $10^{-13}$, respectively. As another test of our results, we have also verified that they are largely insensitive to the interpolation scheme used for $A$. Indeed, the relative differences in the numerical solutions for $\Lambda(r)$ are at most $2\%$ over all $r$ for all the interpolation schemes we have tried.\footnote{\emph{Mathematica} \cite{mathematica} has Hermite and Spline options for interpolation. We have tried both and have also looked at different interpolation orders.} The extracted asymptotic charge $\alpha_4=\sigma_2$ is also highly insensitive to the $A$-interpolation, fractionally changing by at most $10^{-5}$ for the different interpolation methods. 

As a final check, our numerical solutions are also compared with perturbative solutions to the field equations valid approximately near $r = 0$. The solution to Eq.~\eqref{eqn:Asmallcoupling} at small radii is given by
\begin{align}
&A(r)= \sqrt{-\frac{1}{f(r)}} \times\notag \\&\quad\exp\left\{ \bar{\epsilon}\, a(r) \sin\left[\phi_A(r)+ {\cal O}(\tilde{r}^2,\tilde{r}\bar{\epsilon}^2 )\right] \right\}+{\cal O}(\bar{\epsilon})^5 \label{smallrA}
\end{align}
where
\begin{align}
&a(r)= 1 + \frac49 \tilde{r} + {\cal O}(\tilde{r}^2, \tilde{r} \bar{\epsilon}^2)  \\
&\phi_A(r)=F_A(r)+\frac{2 - 3 s_0^2}{24} \sin[2 F_A(r)] \bar{\epsilon}^2 \\
&F_A(r)=\left(1 + \frac{s_0^2\bar{\epsilon}^2}{4}\right) \omega_A \log \tilde{r}  + \phi_0 + \frac{\omega_A}{9} \tilde{r},
\end{align}
$\tilde{r}=r/r_0$, $\omega_A=\sqrt{9 s_0^2-1}/2$, and $\{\bar{\epsilon}, \phi_0\}$ are dimensionless integration constants. Therefore, as $r \to 0$, we expect an oscillatory behavior in $A(r)$ with a steadily decreasing amplitude. 
Note that for $\bar{\epsilon}=0$, the solution for $A$ reduces to a perfectly static aether~\cite{Eling:2006df,Eling:2007xh} (recall that for $r<r_0$, the radial coordinate becomes timelike). 

With this small-$r$ solution for $A(r)$, one can also derive a corresponding asymptotic solution for $\Lambda(r)$. From Eq.~\eqref{eqn:ueqnother}, we get  
\begin{align}
&\Lambda(r)=  r\Lambda_0(r)-\frac{\bar{\epsilon} \kappa}{6r_0 \sqrt{\tilde{r}}} \Lambda_1(r)+{\cal O}(\bar{\epsilon})^4 \label{smallrL}
\end{align}
with
\begin{align}
&\Lambda_0(r)=  \ell \left[1 + \frac{-4 s_0^2+2 s_1^2+1}{3-9 s_0^2} \tilde{r} + {\cal O}(\tilde{r}^2,\bar{\epsilon}^2)\right]\times\notag\\ 
&\quad\sin\left[\omega_\Lambda \log \tilde{r} + \psi_0 + 
\frac{s_0^2+4 s_1^2-1}{9 s_0^2-3}\omega_\Lambda \tilde{r}  + {\cal O}(\tilde{r}^2, \bar{\epsilon}^2)\right]\\
&\Lambda_1(r)=  \left[1+\frac{35 s _0^2+8 s_1^2-8}{18 s_0^2} \tilde{r}
+ {\cal O}(\tilde{r}^2,\tilde{r}\bar{\epsilon}^2)\right]\times\notag \\ &\quad\sin\left[\phi_A(r) -\frac{8 ({s_1^2}-1)}{9 s_0^2} {\omega_A} \tilde{r} + {\cal O}(\tilde{r}^2,\tilde{r}\bar{\epsilon}^2)\right]\notag\\&\quad+\frac16 \bar{\epsilon}^2 \sin^3\left[F_A(r)+{\cal O} (\tilde{r},\bar{\epsilon}^2)\right]
\end{align}
where $\omega_\Lambda=\sqrt{9 s_0^2-4}/2$, and $\{\ell,\psi_0\}$ are again dimensionless integration constants. 
From this, we see that as $r \to 0$, $\Lambda(r)$ also oscillates, but with an amplitude that diverges as $\sim 1/\sqrt{\tilde{r}}$. 
A comparison of this approximate solution with our numerical results is presented in the next subsection.

\subsubsection*{Results}

\begin{figure}[!t]
\begin{center}
 \includegraphics[width=8.5cm]{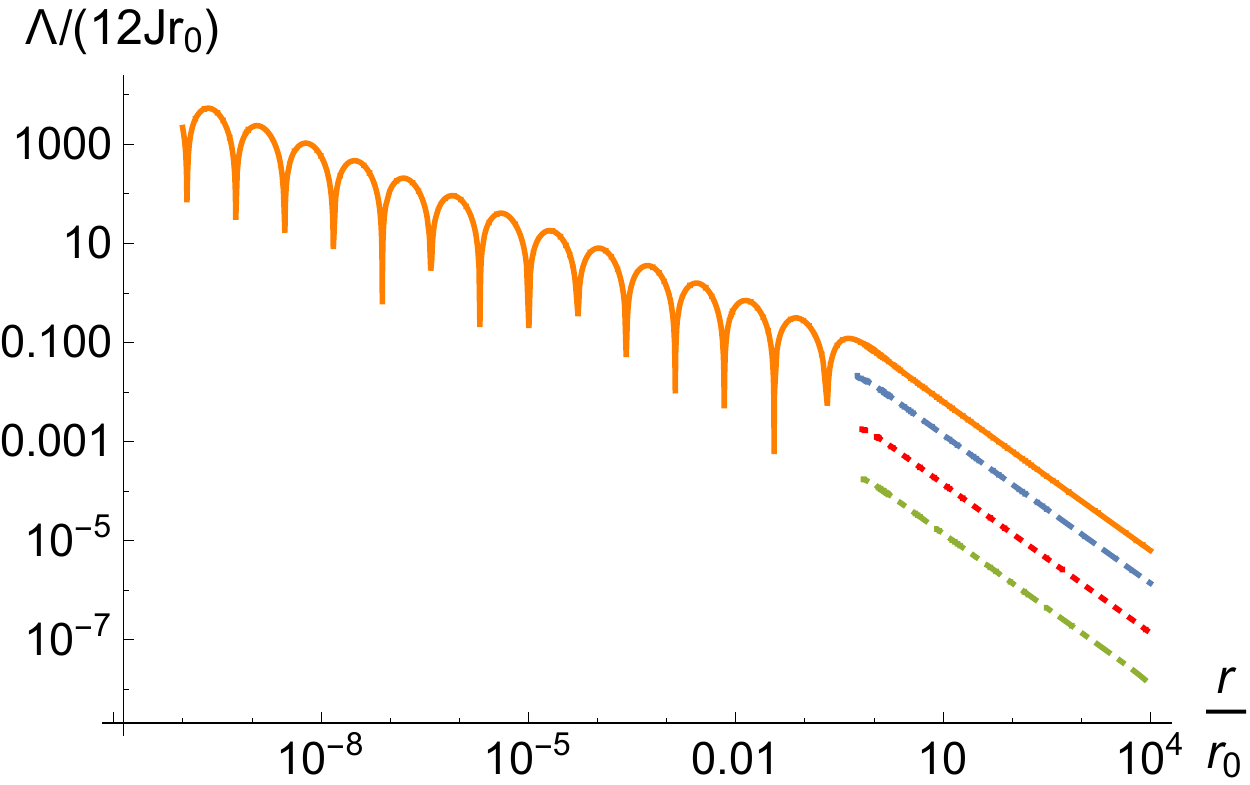}
 \caption{(Color online) Plots of $\Lambda$/(12$J$) for different spin-1 mode speeds (solid orange: $s_1^2 =2$; dashed blue: $s_1^2 =10$, dotted red: $s_1^2 =100$; dash-dotted green: $s_1^2 =1000$) and fixed spin-0 mode speed $s_0^2 = 3/2$. The solution for $s_1^2 =2$ can be extended down to arbitrarily short distances from the central singularity at $r=0$, because no multiple spin-1 horizons are present. The other solutions, however, 
present multiple spin-1 horizons, all of which are singular with the exception of the outermost one. Hence, we only plot those solutions
outside the outermost spin-1 horizon.
}
\label{fig:LambdavsSpin1ModeSpeed}
\end{center}
\end{figure}

Typical results of our numerical integration are displayed in Fig. \ref{fig:LambdavsSpin1ModeSpeed},
which shows the solutions for $\Lambda$ for different values of the spin-1 speed $s_1$.
As can be seen, for small values of $s_1$ the solutions extends to arbitrarily small distances
from the central singularity at $r=0$, which is approached with an oscillatory behavior.
However, as the spin-1 speed is increased (while keeping the spin-0 speed fixed), multiple spin-1 horizons appear. As discussed
previously, regularity can only be imposed at the outermost of these horizons, while finite-area curvature singularities appear at the inner ones.
While phenomenologically acceptable (as these singularities are cloaked by the outermost spin-1 horizon, as well as by the spin-0, spin-2, metric and universal horizons),
this fact prevents us from integrating our solutions down to $r=0$. This can be seen in Fig. \ref{fig:LambdavsSpin1ModeSpeed}, where the solutions corresponding
to $s_1^2=10,$ 100 and 1000 are truncated at a finite radius just outside the finite-area singularity at the second spin-1 horizon.
 
Another key observation to draw from this figure is that as the spin-1 mode speed increases, the \ae ther appears to approach a configuration in which $\Lambda =0$.
(Note that the limit $s_1^2\to \infty$ can be approached \textit{within} the small-coupling approximation.)
This is a hypersurface-orthogonal (in fact, spherically symmetric) configuration.
Now, since $s_1^2\to \infty$ as $c_\omega \rightarrow \infty$ (for generic values of the other couplings), it is tempting to conclude from these results that \ae-theory solutions converge to Ho\v{r}ava-gravity solutions in the infinite-$c_\omega$ limit. However, such a conclusion would be unwarranted because \textit{(i)} one needs to be careful
about how fast $\Lambda$ (and therefore the vorticity) go to zero (c.f. discussion in Sec~\ref{sec:limit}); \textit{(ii)} large $c_\omega$ are incompatible with the small-coupling approximation that we  are using in this section; and \textit{(iii)} for sufficiently large but finite $s_1^2$, $\Lambda(r)$ can be made arbitrarily small at any radius $r$ outside the second spin-1 horizon, but the solution
will always be singular there (i.e. $\Lambda$ diverges at the second spin-1 horizon). We will return to this in  Sec.~\ref{sec:mattingly_solns1}, where we will present an explicit
example of the convergence of \ae-theory solutions to Ho\v rava gravity ones as $c_\omega\to\infty$, and we will discuss these issues in greater detail.

It is noteworthy, though, that it is the regularity condition at the outermost spin-1 horizon that forces the \ae ther into a hypersurface-orthogonal configuration as $s_1^2\to\infty$. Without this regularity condition, Eq.~(\ref{eqn:ueqnother})  can have a wide variety of non-hypersurface-orthogonal solutions even as $s_1^2 \rightarrow \infty$. 
Indeed, for $s_1^2 \rightarrow \infty$, 
the spin-1 regularity condition, Eq.~\eqref{reg}, becomes
\begin{equation}
	\lim_{s_1^2 \rightarrow \infty} \Lambda(r_s) \propto (1+f(r_s)A(r_s)^2) \Lambda'(r_s) \rightarrow 0.
\end{equation}
The first factor in the right-hand side above vanishes because the location of the spin-1 horizon converges to that of the (background) universal horizon as $s_1^2 \rightarrow \infty$. Thus, together with the asymptotic boundary condition $\Lambda(\infty) = 0$, the regularity condition $\Lambda(r_s) \to 0$ selects the unique trivial solution $\Lambda(r)=0$  in the limit $s_1^2 \rightarrow \infty$.\footnote{The proof 
that this boundary value problem has a unique solution follows the same logic as for Eq.~\eqref{masterEq}, if one notes that
$h_3/S\to0$ as $s_1\to\infty$ and that $\bar{h}_2/S>0$ outside the outermost spin-1 horizon (once the spherically symmetric static solutions for $f,A,B$ are used).}

\begin{figure}
\begin{center}
 \includegraphics[width=8.5cm]{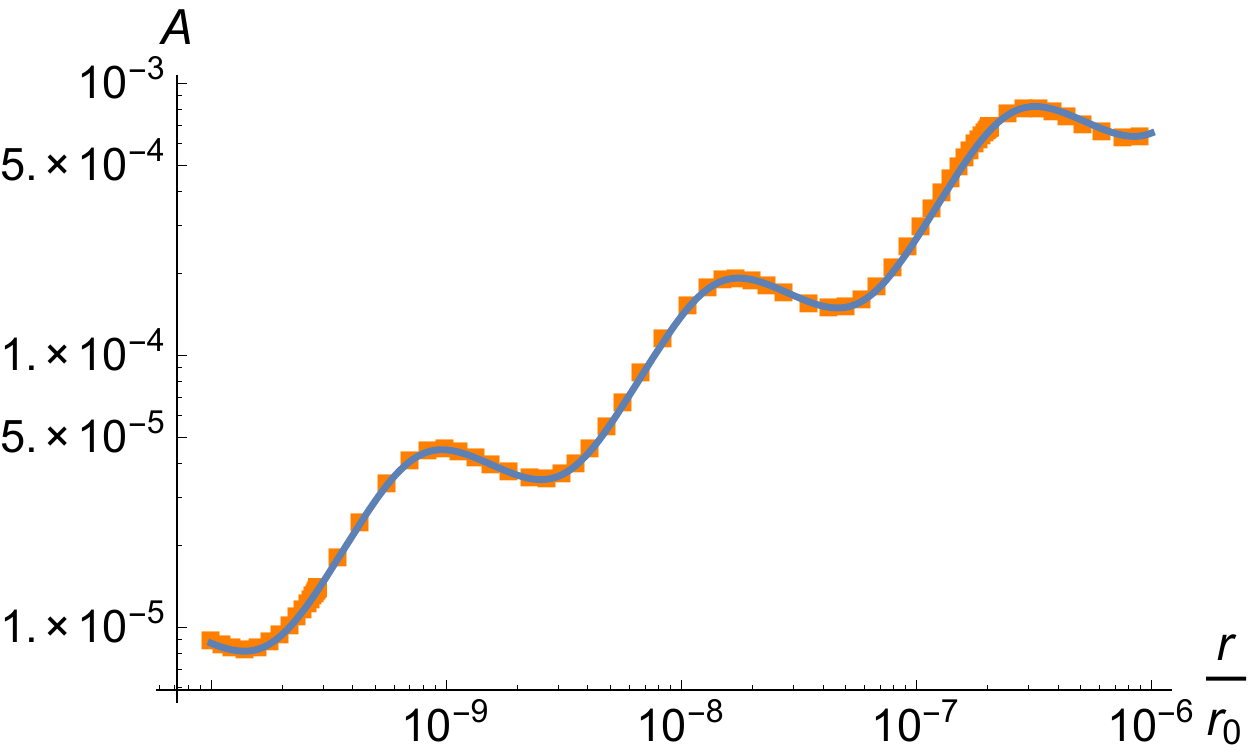}
 \caption{(Color online) Comparing the numerical solution for $A(r)$ (in the small-coupling limit and for $s_0^2 \approx 1.87$ and $s_1^2 \approx 1.95$; orange dots) to a perturbative approximate solution valid at 
sufficiently small radii (solid blue curve).
\label{fig:AsympVsNumA}}
\end{center}
\end{figure}

\begin{figure}
\begin{center}
 \includegraphics[width=8.5cm=8cm]{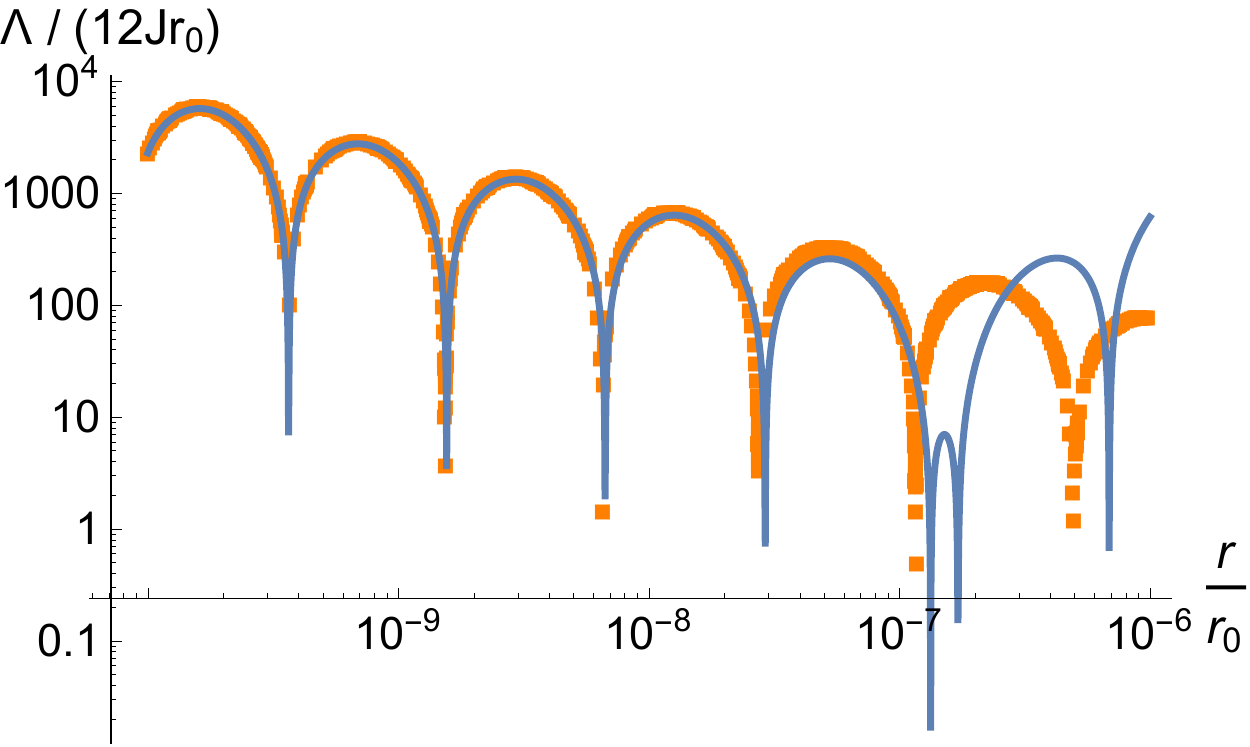}
 \caption{
(Color online) Same as in Fig.~\ref{fig:AsympVsNumA}, but for  $\Lambda(r)$.
}
\label{fig:AsympVsNumLambda}
\end{center}
\end{figure}

In Figs.~\ref{fig:AsympVsNumA} and \ref{fig:AsympVsNumLambda}, we demonstrate the agreement between our numerical solutions and the perturbative solutions given 
in Eq.~\eqref{smallrA} and Eq.~\eqref{smallrL}, which are approximately valid at small radii. The perturbative solution for $A(r)$ depends on two dimensionless constants, $\{\bar{\epsilon}, \phi_0\}$, which are determined by fitting to the numerical data. Figure~\ref{fig:AsympVsNumA} compares this fit to our numerical results. The best-fit values for $\{\bar{\epsilon}, \phi_0\}$ are then used as input for the perturbative solution for $\Lambda(r)$, given by Eq.~\eqref{smallrL}. This solution still depends on another pair of dimensionless constants, $\{\ell,\psi_0\}$, which are also determined by fitting to the data. 
This fit is compared to our numerical solutions in Fig.~\ref{fig:AsympVsNumLambda}.

\begin{figure}[!b]
\begin{center}
 \includegraphics[width=8.5cm]{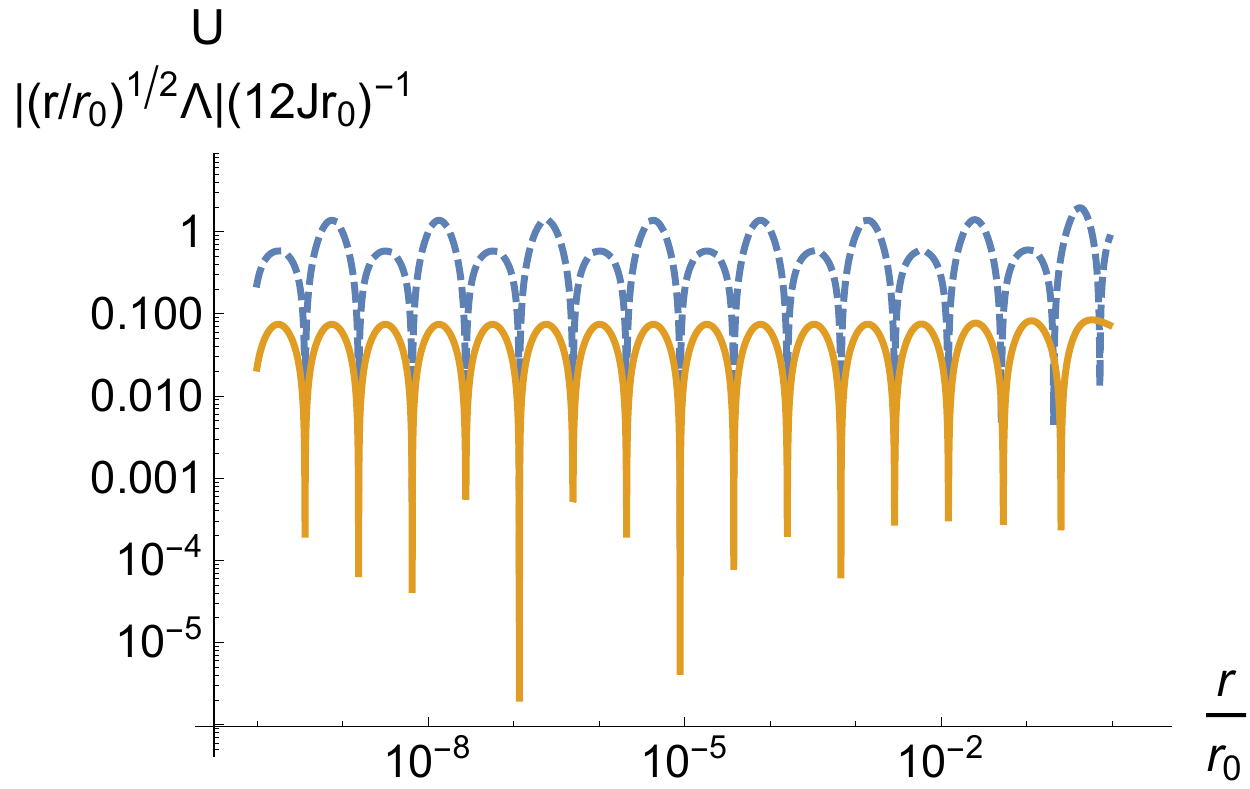}
 \caption{(Color online). Numerical solutions for $U$ (dashed blue) and $r^{1/2}\Lambda$ (solid orange) in the small-coupling limit, and for $s_0^2 \approx 1.87$ and $s_1^2 \approx 1.95$.}
\label{fig:LambdaU}
\end{center}
\end{figure}

Our solutions can also be used to check explicitly whether the necessary condition for the existence of universal horizons can be satisfied in \ae-theory when one switches on
slow rotation. As discussed in Sec.~\ref{sec:uhs}, it is sufficient to verify whether there are any
locations $r=r_u$ such that $\Lambda(r_u)=U(r_u)=0$ or  $\Lambda(r_u)=\Lambda'(r_u)=0$. The typical behavior
of our solutions is displayed in Fig. \ref{fig:LambdaU}. Clearly, $\Lambda$ and $\Lambda'$ never vanish at the same location,
but as the radial coordinate gets smaller, the zeros of $\Lambda$ and $U$ appear to converge.
Since it is numerically challenging to determine whether  $\Lambda$ and $U$ vanish exactly at the same radius
when $r$ is small, we resort to the approximate analytical solutions Eqs.~\eqref{smallrA}--\eqref{smallrL}. 
Those solutions show that $U=0$ if and only if $\phi_A=F_A=0$, and therefore the zeros of $U$ never coincide exactly with those of $\Lambda$. This is
the case even if the integration constant $\ell$ is set to zero. Indeed, if $\ell=0$ the zeros of
$U$ coincide with those of $\Lambda$ only in the limit $r\to 0$, when the terms of ${\cal O}(r)$ in the arguments of the 
oscillatory functions appearing in Eqs.~\eqref{smallrA} and~\eqref{smallrL} can be neglected. Hence, we can conclude that
there are no universal horizons for the slowly rotating solution in the small-coupling limit.

\subsection{Solutions beyond the small-coupling limit}
\label{sec:gen_soln}
We now go beyond the small-coupling limit and solve the full field equations, Eqs.~\eqref{eqns:new}. As mentioned above, 
once all known experimental constraints are taken into account, the allowed part of the parameter space is rather limited. For this reason, 
the dependence of the solutions on the coupling constants is weak, and for presentation
purposes we focus on one special choice, namely
$c_\theta = c_a \approx -0.00305, c_\sigma = 0.01, c_\omega = 0.0018$. The solution for this choice of the coupling constants shares the same qualitative features as the solutions for other viable $c_i$. 
Also note that these coupling constants correspond to $s_0^2 \approx 1.87$ and $s_1^2 \approx 1.95$, 
and that they are sufficiently small to warrant a comparison with the corresponding small-coupling solution. The static, spherically symmetric solution for these values of the coupling  constants that we use as a ``seed'' is obtained by following Ref.~\cite{Barausse-etal-PRD:2011}.

Unlike in the small-coupling case, three pieces of initial data, $\{\Lambda,\Lambda',\Omega'\}$, are needed to fully specify a solution of Eq.~(\ref{eqns:new}). Nevertheless, we can proceed in a manner similar to the small-coupling case discussed earlier. As before, one needs to find initial conditions that correspond to regular, asymptotically flat, slowly rotating solutions. Rescaling our radial coordinate, we first set the location of the metric horizon, $r_0$, to 1 without loss of generality. We then take advantage of the homogeneity of Eq.~(\ref{eqns:new}), which implies that for any constant $K$, $(K \Lambda,K\Omega')$ is a solution to Eq. (\ref{eqns:new}) if $(\Lambda,\Omega')$ is already a solution. We are thus free to set $\Omega'$ at the spin-1 horizon to 1: $\Omega'(r_s)=1$.~\footnote{Note that a rescaling $\Omega'(r)$ was also performed in the small-coupling case, when the integration constant $\kappa$ in Eq. (\ref{eqn:oueqns}) was set to 1.} In doing so, we end up having to deal with a problem similar to the one encountered in the small-coupling case, because  now only $\{\Lambda(r_s),\Lambda'(r_s)\}$ are needed to specify a solution.

Regularity at the spin-1 horizon is guaranteed by imposing either of the equivalent conditions Eq.~(\ref{eqn:reg1}) or Eq.~(\ref{eqn:reg2}). Like before, this means that starting an integration of Eq. (\ref{eqns:new}) requires solely $\Lambda(r_s)$ as input. A given choice of $\Lambda(r_s)$ fixes $\Lambda'(r_s)$, but as shown in Eq. (\ref{eq:uasymptotics}) this will generically lead to divergent solutions as $r\rightarrow\infty$. We thus wish to find solutions for which $\sigma_3 =0$, so that they do not diverge
and are asymptotically flat.
This is again done by a shooting method, as discussed in the previous section. We use $|\sigma_3| < 10^{-12}$ as a stopping condition, with $\sigma_3$ extracted by fitting the numerical result at large radii $r>r_\infty$ to the asymptotic solution in Eqs.~(\ref{eq:oasymptotics}) and~(\ref{eq:uasymptotics}). We use $r_\infty = 1000$ for all our results, but we have verified that they are robust against this choice (e.g. doubling our choice for $r_\infty$ induces fractional changes on the final $\Lambda(r_s)$ of $<10^{-6}$).
Also, as in the small-coupling case, we have verified that the results are robust against the interpolation of $A$.

\begin{figure}[!t]
\begin{center}
 \includegraphics[width=8.5cm]{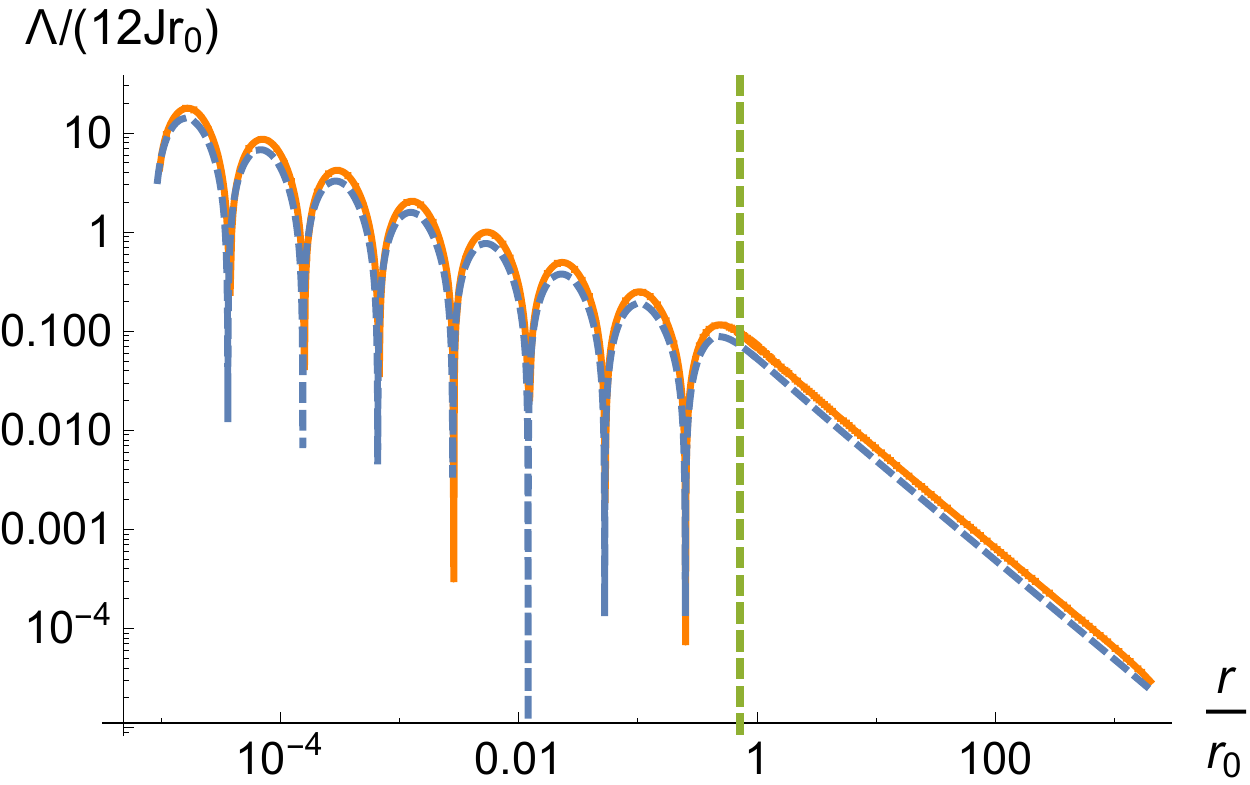}
 \caption{(Color online) 
Solutions for $\Lambda(r)$ for finite values of the coupling constants ($c_\theta = c_a \approx -0.00305, c_\sigma = 0.01, c_\omega = 0.0018$; dashed blue),
together with the corresponding 
small coupling limit solution (solid orange; with $s_0^2 \approx 1.87$ and $s_1^2 \approx 1.95$). The green vertical line marks the universal horizon of the static spherically symmetric seed solution.}\label{fig:NonDecVsDec}
\end{center}
\end{figure}

\begin{figure}[!t]
\begin{center}
 \includegraphics[width=8.5cm]{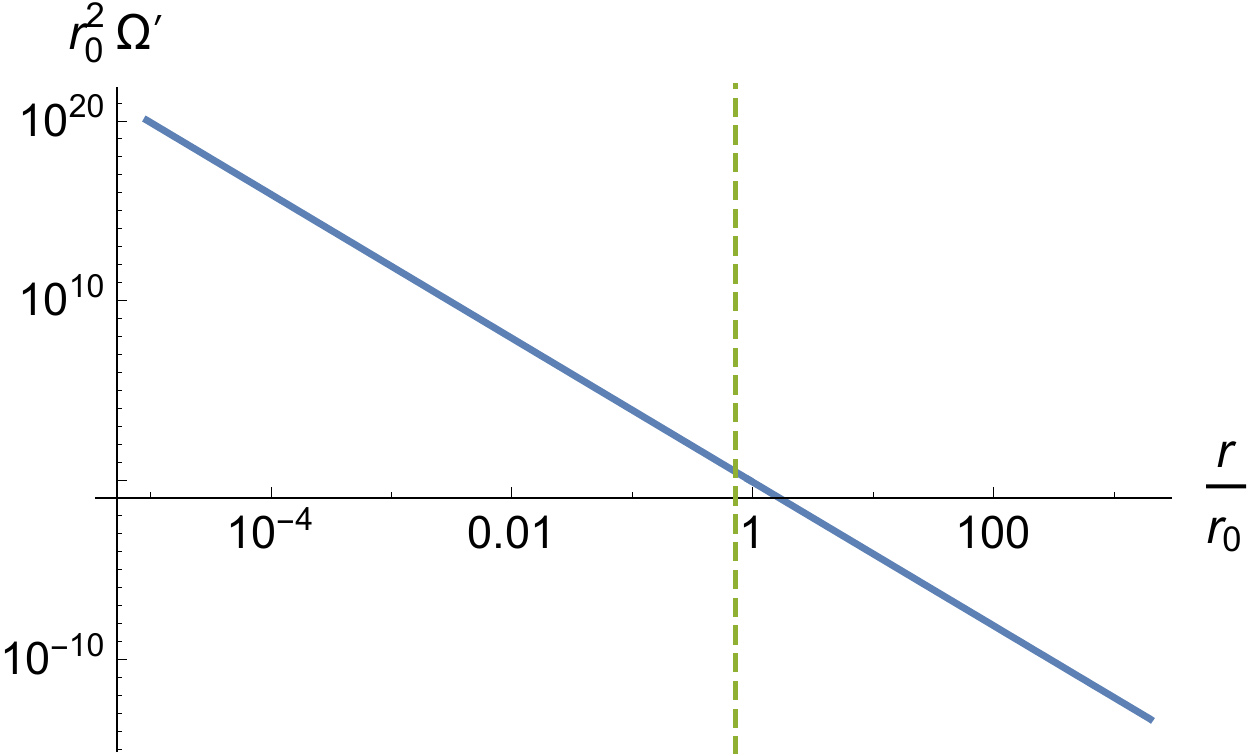}
 \caption{(Color online) Plot (log-log scale) of our numerical solution for $\Omega'(r)$ for $c_\theta = c_a \approx -0.00305, c_\sigma = 0.01, c_\omega = 0.0018$. 
 The green vertical line marks the universal horizon of the static spherically symmetric seed solution. Note that unlike in Fig.~\ref{fig:NonDecVsDec}, we have not
shown the corresponding small coupling limit solution because it would be indistinguishable by eye.
\label{fig:OmegaPrime}}
\end{center}
\end{figure}

Once the desired value of $\Lambda(r_s)$ is found, we again use this initial condition to integrate inward down to very short distances from the central singularity at $r=0$. In Figs.~\ref{fig:NonDecVsDec} and \ref{fig:OmegaPrime} we display $\Lambda$ and $\Omega'$, which we recall are related to the azimuthal component of the aether and the $g_{t\phi}$ component of the metric, respectively.
Not surprisingly, the aether behaves qualitatively in the same manner as in the small-coupling case. It displays a $1/r$-falloff as $r\rightarrow \infty$, as required by asymptotic  flatness, and an oscillatory behavior as $r\rightarrow 0$. Moreover, the frame-dragging presents a strong $1/r^4$-scaling for all $r$, even well inside the black hole.
As for the possible presence of universal horizons,
$\Lambda$ and $\Lambda'$ never vanish at the same location,
but again the zeros of $\Lambda$ and $U$ get closer and closer as $r\to 0$. However, as discussed in the small-coupling limit,
in general they coincide exactly only for $r\to 0$ [c.f. again the approximate solutions given by 
Eqs.~\eqref{smallrA} and~\eqref{smallrL}]. Hence, it seems that universal horizons do not exist, even away from the small-coupling limit.

\begin{figure}[!t]
\begin{center}
 \includegraphics[width=8.5cm]{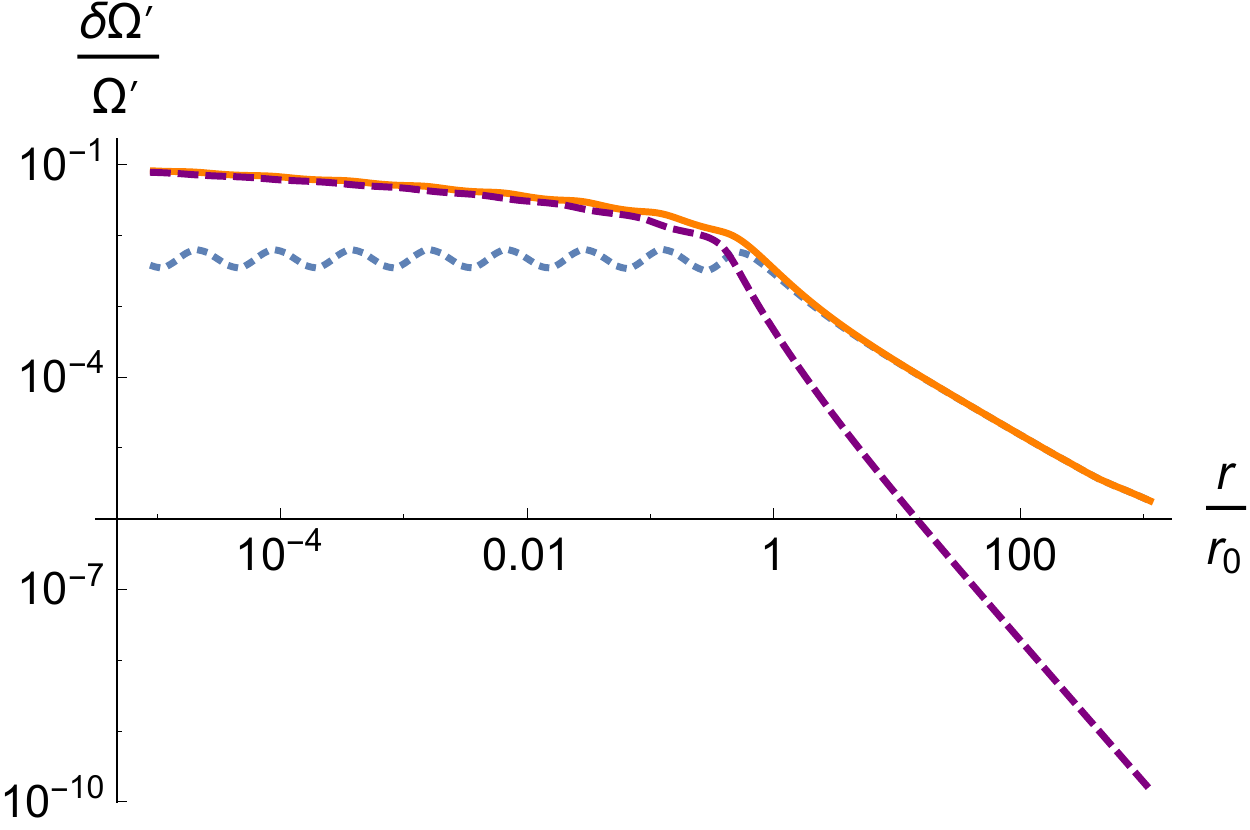}
 \caption{(Color online) Fractional differences (in log-log scale) between the frame-dragging, $\Omega'$, in \ae-theory/Ho\v rava gravity (for $c_\theta = c_a \approx -0.00305, c_\sigma = 0.01, c_\omega = 0.0018$) and GR. In dotted blue is the difference between \ae-theory and Ho\v rava gravity; in solid orange is the difference between \ae-theory and GR; and in dashed purple the difference between Ho\v rava gravity and GR.}
\label{fig:OmegaPrimeComp}
\end{center}
\end{figure}

Each of \ae-theory, Ho\v rava gravity, and GR possesses a two-parameter family of asymptotically flat, slowly rotating black hole solutions, the two parameters being the mass and spin. A 
direct comparison is therefore straightforward and is shown in Fig.~\ref{fig:OmegaPrimeComp},
where we present the differences between \ae-theory and Ho\v rava gravity (for $c_\theta = c_a \approx -0.00305, c_\sigma = 0.01, c_\omega = 0.0018$) and GR.
We recall that in GR, $\Omega'_{\rm GR}/(12J) = 1/r^4$. 
In Ho\v rava gravity, this becomes $\Omega'_{\rm Hor}/(12J) = B(r)/r^4$, where $B(r) \rightarrow 1+O(1/r^2)$ as $r\rightarrow \infty$ \cite{Barausse:2013nwa}.  Equation~\eqref{eq:oasymptotics} instead implies that in \ae-theory $\Omega'_{\rm \ae}/(12J) = 1/r^4 +O(1/r^5)$ asymptotically. We thus expect the fractional differences between \ae-theory and GR/Ho\v rava gravity to fall as $\sim 1/r$ as $r \to \infty$, while the fractional difference between GR and Ho\v rava gravity should fall like $1/r^2$.
This is indeed reflected in Fig.~\ref{fig:OmegaPrimeComp}, which also highlights that  
the differences away from GR remain below percent level throughout the exterior of the black hole.

\subsection{Solutions for $c_\theta + 2 c_\sigma=0$}
\label{sec:mattingly_solns1}

As mentioned previously, an exact static, spherically symmetric solution has been found in Ref.~\cite{Berglund:2012bu} for a 
special combination of the coupling constants that sets the spin-0 propagation speed to zero, i.e.   $c_\theta + 2 c_\sigma=0$.
Below we will use this solution as a ``seed'' to derive slowly rotating black holes, and  study explicitly their limit  as $c_\omega\to \infty$, in which they should become Ho\v rava gravity black holes (c.f. Sec.~\ref{sec:limit}).
The solution found by Ref.~\cite{Berglund:2012bu} is given explicitly by
\begin{align}
	ds^2 &= f dt^2 - \frac{B^2}{f} dr^2 - r^2 d\Omega^2, \\  
	u_\alpha dx^\alpha &= \frac{1+f(r)A(r)^2}{2A(r)}dt + \frac{B(r)}{2A(r)}\left[\frac{1}{f(r)}-A(r)^2\right]dr
\end{align}
where
\begin{align}
	f = 1 - \frac{2\mu}{r} - \frac{\mathcal{R}(2\mu+\mathcal{R})}{r^2}, \,\,\,\,\, B = 1\,,  \label{eq:B}
\end{align}
\begin{equation}
	A = \left(1+\frac{\mathcal{R}}{r}\right)^{-1},
\end{equation}
and $\mathcal{R}$ is a constant given by
\begin{equation}
	\mathcal{R} = \mu \left(\sqrt{\frac{2-c_a}{2(1-c_\sigma)}}-1\right)\,.
\end{equation}
Note that for the \ae ther to be regular everywhere outside the central singularity at $r=0$,
one must have $c_a \leq 2 c_\sigma$. Also, by requiring $s_2^2 > 0$ one obtains $c_\sigma <1$,
while from $s_1^2>0$ one finds $c_a>0$, provided that $c_\omega>-c_\sigma/(1-c_\sigma)$. (Note that this latter condition
does not follow from any theoretical or experimental bounds, but is justified since our goal is to study the limit $c_\omega\to +\infty$.)
Together, these conditions restrict the solution to the parameter region considered by Ref.~\cite{Berglund:2012bu}, i.e.
\begin{equation}
	0 < c_a \leq 2 c_\sigma < 2.
\end{equation}
Moreover, let us note 
that for this family of solutions, the universal horizon is located at $r_u = \mu$, while the spin-0 horizon is effectively pushed to
infinity since the spin-0 propagation speed vanishes.\footnote{Note that even though the spin-0 horizon is pushed to infinity, one still needs
to impose regularity on it~\cite{Berglund:2012bu}, just as one does when it lies at finite radii~\cite{Barausse-etal-PRD:2011}. See footnote 25 in Ref.~\cite{Berglund:2012bu} for more details.}

In order to find the slowly rotating counterparts to these spherically symmetric, static solutions, we need to solve Eq.~(\ref{eqns:new}).
The coefficients appearing in those equations are given, near the universal horizon (where $U=0$),
by
\begin{align}
	p_1 &= 32 + O(U) \\
	p_2 &= 128\left(\frac{c_\sigma - c_a}{1-c_\sigma}\right) + O(U) \\
	p_3/U &= - 256\left(\frac{c_\sigma}{1- c_\sigma}\right) + O(U)
\end{align}
and
\begin{align}
	q_1 =\,\,&2\left(\frac{1-c_\sigma}{2-c_a}\right) + O(U) \\
	q_2 = &-16 + O(U) \\
	q_3/U = &\,\,32\left(\frac{c_a(c_\sigma-2)+2(c_\sigma+c_\omega-c_\sigma c_\omega)}{(2-c_a)c_a}\right) + O(U).
\end{align}
Hence, as previously mentioned, Eq.~(\ref{eqns:new}) 
is regular at the universal horizon.

Let us now consider the spin-1 horizons. In general, this solution has actually two such horizons, which lie at
\begin{equation}
	r_s=\mu\,\pm \,\frac{\mu}{s_1}\,\sqrt{\frac{2-c_a}{2(1-c_\sigma)}}. 
	\label{eqn:spin1loc}
\end{equation}
As $c_\omega \rightarrow \infty$, $s_1 \to \infty$ and $r_s \rightarrow \mu$, i.e. both spin-1 horizons approach the universal horizon as the spin-1 propagation speed diverges. 
Also, in the opposite limit, one can verify that the outer spin-1 horizon is pushed to infinite radius when the propagation speed of the spin-1 mode vanishes, whereas the inner one disappears
in that limit. In fact, the inner horizon ceases to exist when $s_1^2\leq (2 - c_a)/[2(1-c_\sigma)]$.
We also note that we can impose regularity only on one of the two spin-1 horizons, as already discussed in Sec.~\ref{bound}. As a result, if we impose regularity on the outer horizon, the inner spin-1 horizon will be singular. In Appendix \ref{app:manyspin1} we discuss this in more detail, and exclude the possibility that the inner horizon may be ``accidentally'' regular.

The other boundary condition to impose on Eq.~(\ref{eqns:new}) is asymptotic flatness. Because of the vanishing spin-0 propagation velocity, 
the solution near spatial infinity for the special combination 
$c_\theta + 2 c_\sigma=0$ of the coupling constants considered here differs from Eqs.~(\ref{eq:oasymptotics}) and~(\ref{eq:uasymptotics}) (which are otherwise valid if $s_0\neq0$).
In more detail, 
the difference is inherited from the spherically symmetric solutions. As discussed in Ref.~\cite{Berglund:2012bu} and as mentioned above, 
the vanishing spin-0 graviton speed pushes the spin-0 horizon to spatial infinity. As a result, the regularity of the spin-0 horizon needs to be imposed
there, which 
modifies the asymptotic structure of the solutions.

\begin{figure}[!t]
\begin{center}
 \includegraphics[width=8.5cm]{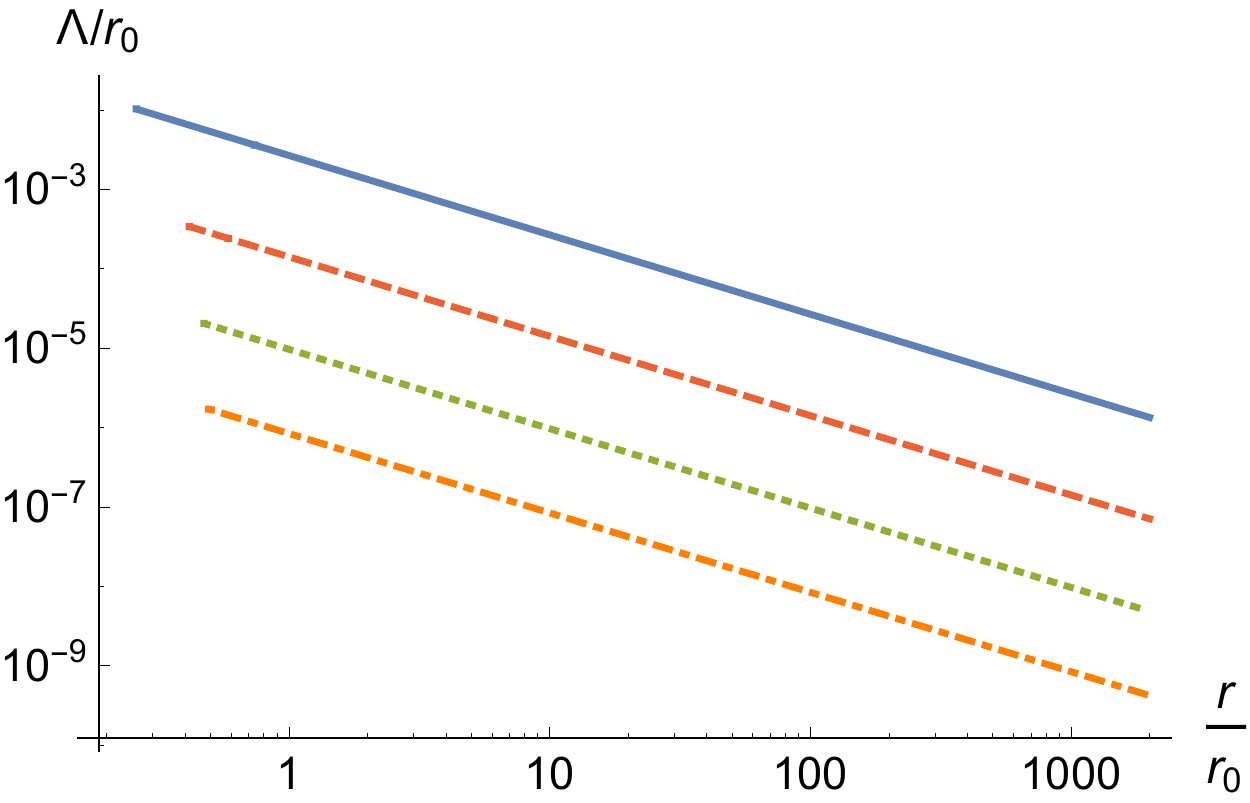}
 \caption{(Color online) $\Lambda(r)$ for selected values of $c_\omega$ (and fixed $c_a=1/2$ and $c_\sigma=3/4$). The solid blue is for $c_\omega=10$, the dashed red one 
for $c_\omega=100$, the dotted green one for $c_\omega=1000$, and the dot-dashed orange one for $c_\omega=10^4$. Note that for each of these cases, a second spin-1 horizon resides within the universal horizon of the spherically symmetric static seed solution. This horizon is singular (i.e. it is the location of a finite-area curvature singularity), and therefore the curves displayed here are terminated right before reaching it. Still, outside this
finite-area singularity,  $\Lambda(r)$ approaches zero at all radii as $c_\omega \rightarrow \infty$.}
\label{fig:UvsOmega1}
\end{center}
\end{figure}

More explicitly, for $c_\theta + 2 c_\sigma=0$ the asymptotic solution reads
\begin{align}
	\Omega' &= \sigma_1 \Omega_1 + \sigma_2 \Omega_2  +\sigma_3 \Omega_3 \\ 
	\Lambda &= \sigma_1 u_1 + \sigma_2 u_2 + \sigma_3 u_3, 
	\label{eq:uasymptoticsSpecial}
\end{align}
where $\sigma_1$, $\sigma_2$ and $\sigma_3$ are integration constants, and
\begin{align}
	\Omega_1 &= \frac{1}{r^4}+\frac{c_a c_\sigma r_0}{(c_\sigma + c_\omega -c_\sigma c_\omega)r^5} + \mathcal{O}\left(\frac{1}{r^6}\right) \,,\\
	u_1 &= \frac{c_a(1-c_\sigma)}{8(c_\sigma + c_\omega -c_\sigma c_\omega)r^2} + \mathcal{O}\left(\frac{1}{r^3}\right)\,,
\end{align}
\begin{align}
	\Omega_2 &= \frac{2c_a (3c_\sigma + c_\omega)}{(c_\sigma+c_\omega - c_\sigma c_\omega)r^5} + \mathcal{O}\left(\frac{1}{r^6}\right) \,,\\
	u_2 &= \frac{1}{r} + \frac{[c_a(2-3c_\sigma)+2c_\sigma(1-c_\omega)+2c_\omega]r_0}{4(c_\sigma + c_\omega -c_\sigma c_\omega)r^2} + \mathcal{O}\left(\frac{1}{r^3}\right)\,,
\end{align}
\begin{align}
	\Omega_3 &= \frac{2c_a c_\omega}{(c_\sigma + c_\omega -c_\sigma c_\omega)r^2} + \mathcal{O}\left(\frac{1}{r^3}\right) \,,\\
	u_3 &= r^2 + \frac{(c_\sigma + c_\omega -c_\sigma c_\omega -2c_a)r_0r}{2(c_\sigma + c_\omega -c_\sigma c_\omega)}+\mathcal{O}\left(r^0\right)\,.
	\label{eq:asympdiv}
\end{align}
Evidently, the last mode is the divergent one. Asymptotically flat slowly rotating black holes are therefore those for which $\sigma_3 =0$. 

The numerical integration of Eq.~(\ref{eqns:new}) is then performed as outlined in Sec.~\ref{sec:gen_soln}. Figure~\ref{fig:UvsOmega1} displays solutions with increasing $c_\omega$ but fixed $c_a=1/2$ and $c_\sigma=3/4$. (Note that this figure represents the generic behavior of solutions for this sector.) 
What can be immediately observed is that Fig.~\ref{fig:UvsOmega1}  closely mimics the behavior of the small-coupling solutions of Fig.~\ref{fig:LambdavsSpin1ModeSpeed} as $s_1\to\infty$.  
As $c_\omega \rightarrow \infty$, $s_1^2$ also diverges -- which explains
the similarity between Figs.~\ref{fig:LambdavsSpin1ModeSpeed} and \ref{fig:UvsOmega1} --, and $\Lambda(r)$ goes to zero at all radii -- which represents a hypersurface-orthogonal (actually, spherically symmetric) \ae ther configuration. 

\begin{figure}[!t]
\begin{center}
 \includegraphics[width=8.5cm]{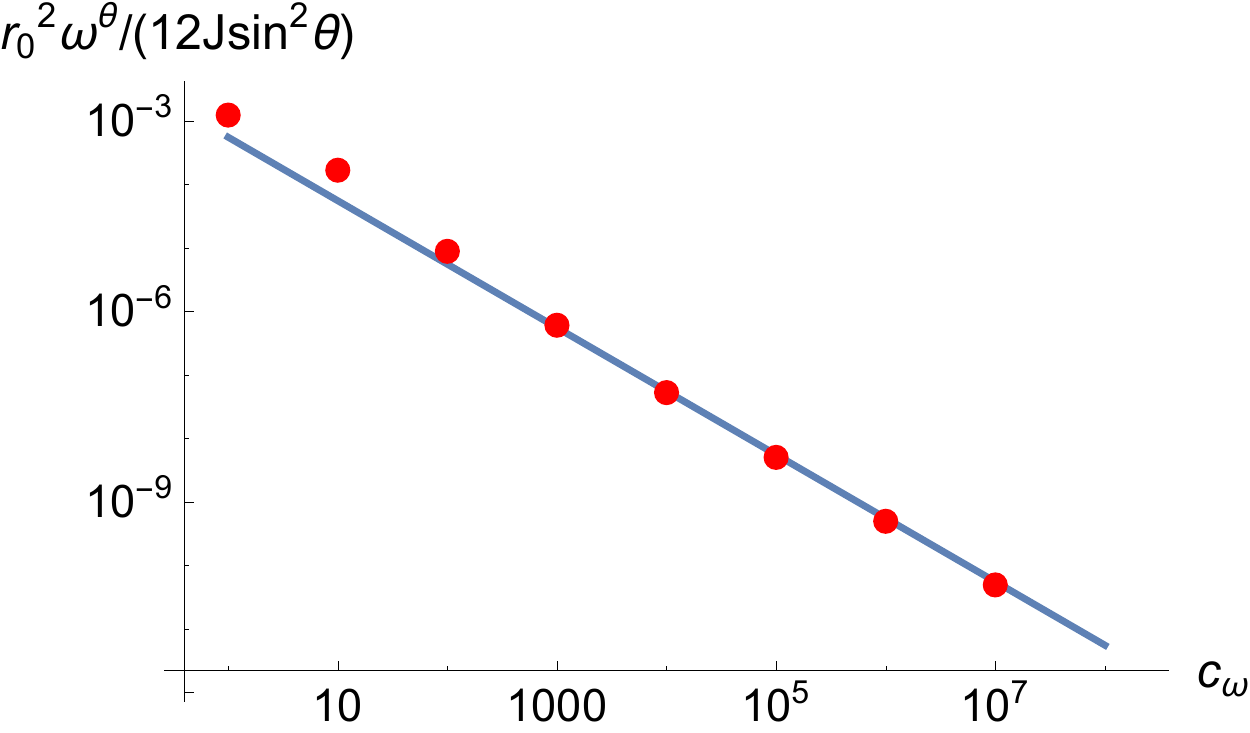}
 \caption{(Color online) Behavior of $\omega^\theta/\sin^2\theta$ evaluated at $r=4\mu$ as a function of $c_\omega$ (for fixed $c_a=1/2$ and $c_\sigma=3/4$) from our numerical solutions (red dots) vs a simple $1/c_\omega$ scaling (solid blue line).}
\label{fig:twistvscw}
\end{center}
\end{figure}

In order to assess whether 
in the limit $c_\omega\to \infty$ the \ae-theory solutions converge to Ho\v rava gravity ones, however, we need 
to look at how fast $\Lambda(r)$ (and therefore the vorticity)  goes to zero (c.f. Sec.~\ref{sec:limit}).
This is shown in Fig.~\ref{fig:twistvscw}, which
compares the value of the $\theta$ component of the twist vector -- evaluated at $r = 4 \mu$ for  $c_a=1/2$, $c_\sigma=3/4$ and several values of $c_\omega$ -- with the $1/c_\omega$ scaling
expected from  Sec.~\ref{sec:limit}. Based on the considerations of that section, the fact that 
$\Lambda(r)$ scales as  $1/c_\omega$ is enough to ensure that the \ae-theory slowly rotating
solutions that we study converge to Ho\v rava gravity solutions as $c_\omega\to \infty$.

\begin{figure}[!t]
\begin{center}
 \includegraphics[width=8.5cm]{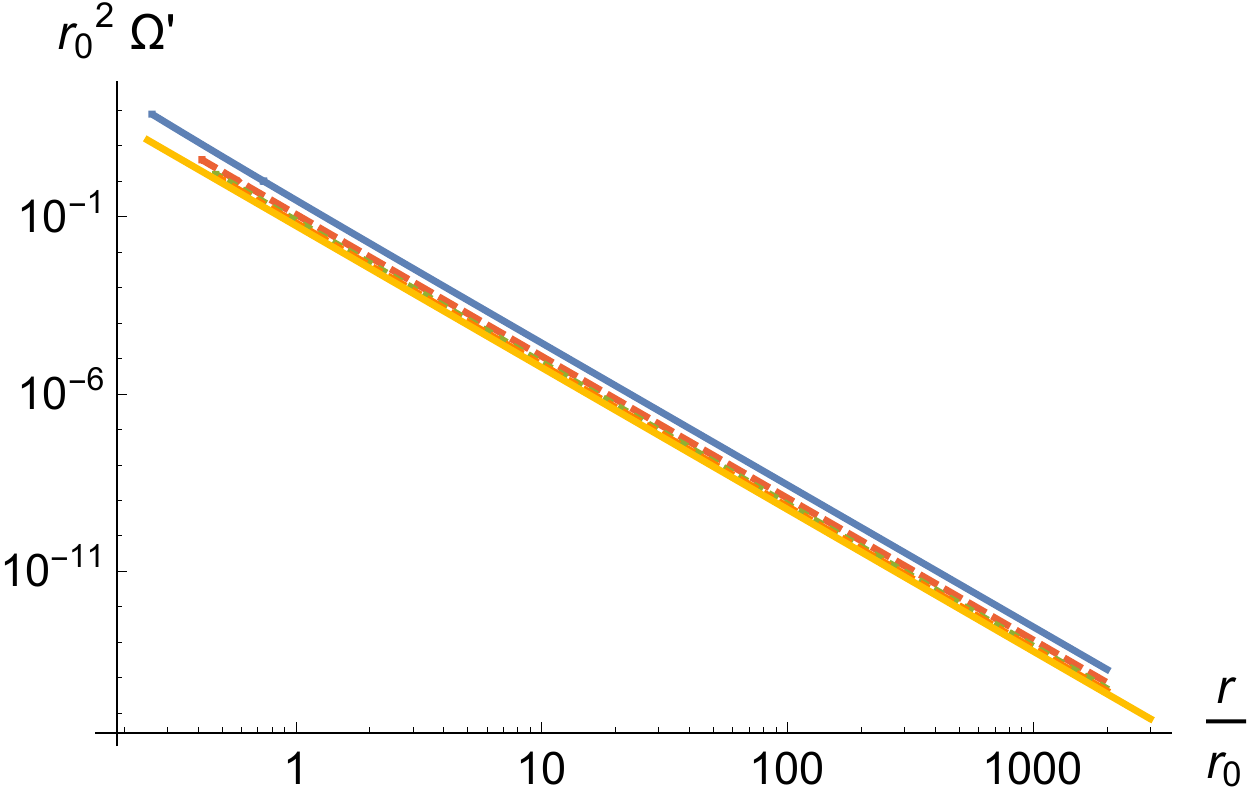}
 \caption{(Color online) Derivative of frame dragging, $\Omega'(r)$, for various values of $c_\omega$ following the same color scheme of Fig.~\ref{fig:UvsOmega1}. The additional yellow curve below all other curves is the GR result, which also coincides with the Ho\v{r}ava gravity result for $c_{a}+2c_\sigma = 0$. We see again that as $c_\omega \rightarrow \infty$, the frame dragging in \ae-theory solutions approaches that of Ho\v{r}ava gravity.}
\label{fig:FrameDraggingvsOmega}
\end{center}
\end{figure}

This fact can also be verified directly by comparing the frame-dragging potential $\Omega(r)$ of our solutions to the 
Ho\^{r}ava gravity frame-dragging~\cite{Barausse:2012qh}   
\begin{equation}\label{rot_soln2}
	\Omega(r)= - 12J \int^r_{r_H} \frac{B{(\rho)}}{\rho^4} + \Omega_0,
\end{equation}
where $J$ and $\Omega_0$ are integration constants. Since 
spherically symmetric \ae-theory solutions are also solutions to Ho\v rava gravity, Eq.~\eqref{eq:B} ensures that Ho\v rava gravity black holes
have $B=1$ for $c_a + 2 c_\sigma=0$, hence
the derivative of the frame dragging in Ho\v rava gravity
matches the Kerr behavior
\begin{equation}
	\Omega'(r)= -\frac{12 J}{r^4}\,.
\end{equation}
This is compared to $\Omega'$ in \ae-theory in Fig.~\ref{fig:FrameDraggingvsOmega}. As can be seen, that
figure confirms that Ho\v rava gravity solutions are recovered in the limit $c_\omega\to \infty$.

A subtle point in this limit and in the comparison of the solutions of the two theories has to do with the singularity of the inner spin-1 horizon (c.f. Figs.~\ref{fig:UvsOmega1} and \ref{fig:FrameDraggingvsOmega}). In Ho\v rava gravity the concept of a spin-1 horizon is absent,  as there is no spin-1 excitation. However, when comparing solutions one can still compare the metric and the aether configuration at the radius of the inner spin-1 horizon in \ae-theory. Since in Ho\v rava gravity there is no correction to the aether's configuration to first order in rotation, there cannot be any singularity at that radius. As $c_\omega\to \infty$, $s_1\to \infty$ and both the spin-1 horizons of the \ae-theory solutions merge onto the universal horizon of the Ho\v rava gravity  solution. As the two spin-1 horizons merge in that limit, the regularity condition on the outer horizon should therefore be sufficient to ensure that the limit does indeed match the Ho\v rava gravity solution. However, it should also be clear that for any arbitrarily large but finite value of $c_\omega$, the Ho\v rava gravity solution will differ significantly from the corresponding \ae-theory solution in the vicinity of the inner (singular) spin-1 horizon. This should be a point of caution regarding the practical use of large $c_\omega$ solutions in \ae-theory as approximate solutions in Ho\v rava gravity. 

It is worth stressing that the appearance of multiple spin-1 horizons (and therefore of curvature singularities at the location
of all but the outermost of them) for large but finite $c_\omega$ is not just a feature of the solutions with 
$c_\theta + 2 c_\sigma=0$ presented in this section, but is also present for general solutions.
This is easy to understand by looking at the spin-1 metric component $g_{tt}^{(1)}=f(r)+(s_1^2-1) u_t(r)^2$ [$f(r)$ and $u_t(r)$ being defined 
by Eqs.~\eqref{metric} and~\eqref{eqn:aether}], which is zero at the spin-1 horizons. 
As $c_\omega$ is increased (while keeping $c_a$, $c_\sigma$ and $c_\theta$ fixed), $f(r)$ and $u_t(r)$ do not change --
because spherically symmetric static solutions have zero vorticity and thus do not depend on $c_\omega$~\cite{Barausse-etal-PRD:2011} --, while
$s_1$ diverges as per Eq.~\eqref{eqn:s1mode}. In the limit of infinite $c_\omega$, the zeros of 
$g_{tt}^{(1)}$ thus match those of $u_t$, which correspond to the location of the universal horizons of the spherically symmetric static solution.
In general, however, spherically symmetric static solutions admit multiple universal horizons~\cite{Barausse-etal-PRD:2011}, hence it is not surprising that
for large but finite $c_\omega$,  $g_{tt}^{(1)}$ 
will have multiple zeros (thus leading to multiple spin-1 horizons). 

In fact, one can show that
\textit{for {each} universal horizon of the spherically symmetric static solution, two spin-1 horizons will appear, for
large but finite $c_\omega$.} To see this, let us first note that $g_{tt}^{(1)}=0$
 implies $u_t(r_s)=\pm \sqrt{-f(r_s)/(s_1^2-1)}=\pm {\cal O}(1/s_1)$ for large $c_\omega$ (and thus large $s_1$), $r_s$ being a spin-1 horizon's location.
From this equation, it also follows that $r_s\to r_u$ as $s_1\to\infty$ (with $r_u$ a universal horizon of the spherical solution). Now, since
$\mbox{d}(u_t)/\mbox{d}r(r_u)\equiv k \neq0$ (c.f. e.g. Figs. 10--12 in Ref.~\cite{Barausse-etal-PRD:2011}), we can write $u_t =k (r-r_u)+{\cal O}(r-r_u)^2$ 
in the vicinity of the universal horizons of the spherical solution. 
Since $r_s\to r_u$ as $s_1\to\infty$, we can use this approximation for $u_t$ when solving $g_{tt}^{(1)}=0$ for $r_s$ [i.e. when solving the equation $u_t(r_s)=\pm \sqrt{-f(r_s)/(s_1^2-1)}=\pm {\cal O}(1/s_1)$]. This yields the two solutions $r_s=r_u\pm {\cal O}(1/s_1)$ for large $s_1$. Therefore, for large but finite $s_1$, there will be two spin-1 horizons for each universal horizon, one on each side of the latter. We have indeed verified this result using the numerical solutions of Ref.~\cite{Barausse-etal-PRD:2011}. 

\subsection{Solutions for $c_a=0$}
\label{sec:cazero}

In Sec.~\ref{sec:uhs}, 
we discussed (as was already noted in Ref.~\cite{Barausse:2012ny})
 that  $c_a = 0$ constitutes a special case in which \ae-theory admits
hypersurface-orthogonal slowly rotating solutions (i.e. ones with $\Lambda=0$).
These solutions are therefore also solutions to Ho\v rava gravity.

An exact static, spherically symmetric solution for $c_a = 0$ has been given in Ref.~\cite{Berglund:2012bu}:
\begin{align}
	ds^2 &= f dt^2 - \frac{B^2}{f} dr^2 - r^2 d\Omega^2, \\  
	u_\alpha dx^\alpha &= \frac{1+f(r)A(r)^2}{2A(r)}dt + \frac{B(r)}{2A(r)}\left[\frac{1}{f(r)}-A(r)^2\right]dr
\end{align}
where
\begin{align}\label{ca_soln}
	f = 1 - \frac{2\mu}{r} - \frac{c_\sigma r_{\ae}^4}{r^4}, \,\,\,\,\, B = 1\,,
\end{align}
\begin{equation}
	A = \frac{1}{f}\left(-\frac{r_{\ae}^2}{r^2}+ \sqrt{f+\frac{r_{\ae}^4}{r^4}}\right), 
\end{equation}
and regularity of the \ae ther everywhere outside the universal horizon requires 
\begin{equation}
	r_{\ae} = \frac{\mu}{2}\left(\frac{27}{1-c_\sigma}\right)^{1/4}.
\end{equation}
This is a one-parameter family of solutions, parametrized by $\mu$, and the universal horizon is located at 
\begin{equation}
r_u = \frac{3}{2}\mu.
\end{equation}

When $c_a=0$, $s_0^2,s_1^2 \to \infty$, so both the spin-0 horizon and the spin-1 horizon coincide with the universal horizon, where $U=0$. This means that regularity needs to be ensured only on the surface where $U= 0$, i.e. at $r=r_u$. When $c_a = 0$, the coefficients of Eqs.~(\ref{eqns:new}) become 
\begin{align}
	p_1/S &= -\frac{4}{r}, \\
	p_2/S &= 0\,, \\
	p_3/(SU) &= -\frac{2048c_\sigma}{9\sqrt{3(1-c_\sigma)}U}+{\cal  O}(1)\,, \mbox{ as } U \rightarrow 0
\end{align}
and 
\begin{align}
	q_1/S =& \,\,0\,,\\
	q_2/S =& \,\,-\frac{16\sqrt{2(1-c_\sigma)}}{\sqrt{3}U} +{\cal  O}(1)\,, \mbox{ as }U \rightarrow 0\\
	q_3/(SU) =& \,\,\frac{128(1-c_\sigma)}{U^2} +{\cal O}(1/U)\,, \mbox{ as }U \rightarrow 0\,,
\end{align}
which confirms that $U=0$ is indeed a genuine singular point of Eqs.~(\ref{eqns:new}). However, since $q_1=0$ and $p_2=0$ (at all radii), the system given by Eqs.~(\ref{eqns:new}) reduces to
\begin{subequations}
\label{eqns:c14}
\begin{align}
\Omega'' &= \frac{1}{S}\left(p_1\Omega' + \frac{p_3}{U} \Lambda\right)\,,\label{Omegaeq}\\
\Lambda'' &= \frac{1}{S}\left(q_2\Lambda' + \frac{q_3}{U} \Lambda\right)\,, \label{Leq}
\end{align}
\end{subequations}
where $\Lambda$ decouples and can be determined by solving Eq.~(\ref{Leq}) alone. 
Indeed, one can verify that Eq.~(\ref{Leq}) reduces \textit{exactly} to Eq.~\eqref{masterEq} (with the right-hand side
remainder ${\cal O}(1/c_\omega)$ set exactly to zero), once one recalls the relation between $\lambda$ and $\Lambda$ (c.f. Sec. \ref{field_eqs}). One can then follow the reasoning of Sec.~\ref{sec:limit} to conclude that regularity at the spin-1/universal horizon and asymptotic flatness select the unique solution $\Lambda=0$.

For $\Lambda=0$, Eq.~(\ref{Omegaeq}) can be then be integrated to give
\begin{equation}
\Omega(r) = \Omega_0 + \frac{4 J}{r^3},
\end{equation}
where $\Omega_0$ and $J$ are integration constants. 
This matches the frame dragging of a slowly rotating Kerr black hole, and also that of
the Ho\v rava gravity solution given by Eq.~\eqref{rot_soln2} [once one recognizes that $B=1$, c.f. Eq.~\eqref{ca_soln}].

To conclude, when $c_a=0$, there exist slowly rotating \ae-theory solutions with a spherically symmetric (and thus hypersurface-orthogonal) aether configuration. Note that this result is not at odds with the proof of Refs.~\cite{Barausse:2012ny,Barausse:2012qh}, which showed that $\Lambda(r) = 0$ (i.e. hypersurface orthogonality) implies $\Omega'(r) = 0$ (i.e. no rotation). Indeed,  Refs.~\cite{Barausse:2012ny,Barausse:2012qh} explicitly pointed out that $c_a=0$ constitutes an exception to the proof.

\section{Discussion and conclusions}

We have studied slowly rotating, asymptotically flat black holes in \ae-theory. Below we summarize and discuss our main results.

We have started by revisiting the relation between slowly rotating solutions in \ae-theory and in Ho\v rava gravity. As already shown in Refs.~\cite{Barausse:2012ny,Barausse:2012qh}, hypersurface-orthogonal \ae-theory solutions cannot support rotation for generic values of the coupling constants.
This implies that, in general, the slowly rotating solutions of Ho\v rava gravity will not be solutions of \ae-theory and vice versa. 
The special  case $c_a=0$ constitutes an exception. We have considered it separately and have shown that for $c_a=0$
the slowly rotating \ae-theory solutions match the Ho\v rava gravity ones. Remarkably, these solutions also 
share the same frame dragging as the slowly rotating Kerr black holes of GR, although their geometry
does not match the Schwarzschild one when rotation is switched off, due to a non-trivial \ae ther configuration in spherical symmetry.

We have also explored in depth the $c_\omega\to \infty$ limit, previously considered in Ref.~\cite{Jacobson:2013xta}. We have uncovered and clarified several subtleties in applying the logic of Ref.~\cite{Jacobson:2013xta} to slowly rotating solutions, and we have argued that suitable boundary conditions are crucial to ensure that \ae-theory solutions converge to Ho\v rava gravity ones in this limit. In order to have a concrete family of explicit solutions that exhibits this convergence, 
we have generated the slowly rotating counterpart of the exact static, spherically symmetric solution found in Ref.~\cite{Berglund:2012bu} for the special choice $c_\theta + 2 c_\sigma=0$, and we have shown that the \ae ther does indeed become hypersurface-orthogonal as $c_\omega$ diverges.

We have also shown that, for generic values of the coupling constants, there exists a three-parameter family of slowly rotating, asymptotically flat black hole solutions in \ae-theory. However, these solutions generally exhibit finite area singularities. Spin-1 perturbations propagate along null geodesics of an effective metric, the spin-1 metric, and the singularities correspond to the location of the Killing horizons of this metric. The outermost of these Killing horizons acts as an event horizon for the spin-1 perturbations, and solutions for which it is regular constitute  a two-parameter subset of the three-parameter family of the general solutions. 
These two parameters can be interpreted as the mass and the angular momentum of the black hole. This implies that slowly rotating, asymptotically flat \ae-theory solutions with regular outermost spin-1 horizons cannot have independent \ae ther charges. Nevertheless, the solution for the \ae ther is non-trivial, and as a result these black holes always have a hair of the ``second kind''. If more than one spin-1 horizon exists, then the inner ones will not be regular. 

We have resorted to a small-coupling approximation to study in detail the configuration of the \ae ther in the interior of the black hole. In this approximation, one essentially solves the \ae ther equation on the background of a slowly rotating Kerr black hole. Viability constraints on \ae-theory imply that the coupling constants are indeed small, 
so one expects the small-coupling approximation to be quite accurate. Our main concern has been to check whether the aether becomes orthogonal to some constant radius surface, because then such a surface would resemble the universal horizon found in spherically symmetric static black holes~\cite{Barausse-etal-PRD:2011,blas-sibiryakov-prd:2011}. We have shown that this is not the case, hence universal horizons do not exist in slowly rotating, asymptotically flat \ae-theory black holes.

Finally we have generated the full solutions for viable values of the coupling constants, 
and we have verified a remarkable agreement with the small-coupling approximation, in line with our expectations. We have calculated the fractional deviations of the frame dragging potential of our solutions  from the  corresponding GR and Ho\v rava-gravity ones. In all cases, the deviations are too small to be detectable with current observations, but are probably within the reach of future gravitational-wave missions (c.f. the 
Evolved Laser Interferometer Space Antenna  -- eLISA --, which will map the geometry of supermassive black holes with $10^{-5}$ fractional accuracy~\cite{Seoane:2013qna}).

It is possible that rapidly rotating black holes might exhibit more appreciable deviations from GR. Since we have worked within the slow-rotation approximation throughout this paper, we are unable to probe this regime. Another promising future direction which could allow one to distinguish between rotating black holes in \ae-theory and GR is to explore the behavior of perturbations. Irrespective of how similar the black hole backgrounds are, perturbations may differ significantly as the theories have different degrees of freedom~\cite{Barausse:2008xv}. The crucial question that deserves future attention is whether this leads to any observable effects.

\acknowledgments
We are grateful to Ted Jacobson for insightful discussions, and for reading a preliminary version of this manuscript. 
The research leading to these results has received funding  from the European Research Council under the European Union's Seventh Framework Programme (FP7/2007-2013) / ERC grant agreement n. 306425 ``Challenging General Relativity'' (to T.P.S). 
We also acknowledge support from
 the European Union's Seventh Framework Programme (FP7/PEOPLE-2011-CIG)
through the  Marie Curie Career Integration Grant GALFORMBHS PCIG11-GA-2012-321608, and
from the H2020-MSCA-RISE-2015 Grant No. StronGrHEP-690904 (to E.B.). T.P.S.  also thanks the Institut d'Astrophysique de Paris and the
ILP LABEX (ANR-10-LABX-63) for hospitality during
a visit supported through the Investissements d'Avenir
programme under reference ANR-11-IDEX-0004-02. I.V. thanks Barry Wardell and Adrian Ottewill for warmly hosting a visit to the University College Dublin, where parts of this work was conducted. I.V. gratefully acknowledges the Royal Irish Academy for a Charlemont Grant that made this visit possible.

\vskip 3cm
\appendix
\section{Coefficients of the differential equations}
\label{app:coeffs}

The coefficients of Eqs. (\ref{eeq1}) and (\ref{eeq2}) are functions of the spherical solutions. Using $s_1^2$ and $s_2^2$ as the squares of the spin-1 and spin-2 mode speeds, respectively, these coefficients are explicitly given by 
\begin{align}
d_1(r) &= c_a \Bigg[\frac{\left(\left(U-2\right) A'+A^3 f'\right)}{4 r^4 A^2 B^2} - \left(\frac{s_1^2}{s_2^2}\right)\frac{U\left(r B'-4 B\right)}{4 r^5 A B^3}\Bigg], \\ 
d_2(r) &=c_a \left(\frac{s_1^2}{s_2^2}\right)\frac{U}{4 r^4 A B^2},\\ 
d_3(r) &= c_a (16r^6A^4B^3)^{-1}\Bigg[\left(\frac{s_1^2}{s_2^2}-1\right)A(U-2)^2UB' \nonumber \\ &+ B \Bigg((U-2)\left[4-4U +\left(3+s_2^2-2\frac{s_1^2}{s_2^2}\left(2+c_a s_2^2\right)\right)\right. \nonumber\\  &\phantom{......}\times \,\, U^2 \Big] A' + A^3 \left[ 4 + 8\left(\frac{s_1^2}{s_2^2}-1\right)U +\Big(3+s_2^2 \right. \nonumber\\ &\phantom{......}\left. -2\frac{s_1^2}{s_2^2} \left(2+c_a s_2^2\right)\Big)U^2\right]f'\Bigg)\Bigg]\\
d_4(r) &=c_a \left(1-\frac{s_1^2}{s_2^2}\right)\frac{\left(U-2 \right)^2 U}{16 r^6 A^3 B^2},
\end{align}
and,
\begin{align}
b_1(r) &=\frac{1}{s_2^2}\left(\frac{r B'-4 B}{2 r B^3},\right)\\
b_2(r) &=-\frac{1}{s_2^2}\left(\frac{1}{2 B^2}\right),\label{b2}\\
b_3(r) &=(8 r^2 s_2^2 A^3 B^3)^{-1}\Bigg[2 B \Bigg(\Big((c_a-2) s_2^2+2\Big)\times \nonumber \\ &\phantom{================}\Big(A^4 f^2-1\Big) A' \nonumber \\ &+A^3 f' \Big(A^2 \Big((c_a-2)
   s_2^2+2\Big) f+(c_a-2) s_2^2-2\Big)\Bigg) \nonumber \\ &+A B' \Bigg(\Big(s_2^2-1\Big) A^4 f^2+2
   \Big(s_2^2+1\Big) A^2 f+ s_2^2-1\Bigg)\Bigg]\\
b_4(r) &=\frac{4-4U - (s_2^2-1)U^2}{8 r^2 s_2^2 A^2 B^2},\label{b4}
\end{align}
where $U=1+fA^2$. Equation (\ref{eqn:s1mode}) shows that $s_1^2$ depends linearly on $c_\omega$, while the other mode speeds do not depend on $c_\omega$ at all. Hence, the coefficients
$b_i$ are independent of $c_\omega$ and $d_i$ are all linear in $c_\omega$.

\newpage
\section{Explicit expressions for $p_i$ and $q_i$.}
\label{pandq}
\begin{widetext}
\begin{align}
p_1 = \,\,&(s_1^2 U^2 - (U-2)^2) \left(\frac{B'}{B} - \frac{4}{r}\right) 
+ \left(\frac{c_\sigma}{c_\sigma-1}\right)\frac{Uq_1}{r^2A} \\
p_2 = \,\,&\frac{2 c_a s_1^2 U^2q_1}{(1-\cs)r^4A^2} + \frac{1}{(\cs-1)^2 r^2 A^2}\Bigg[\frac{q_1}{r^2} \Big(8(\ca-\cs)(\cs-1)(1-U) + (2\ca(\cs -1) -\cs^2)U^2 \Big) -4\cs(\cs-1)Uf'A^3\Bigg] \\ 
p_3 = \,\,&\frac{2s_1^2U^2}{(\cs-1)}\Bigg[\frac{\ca(U-2)^2A'{}^2}{r^2A^3}+\frac{4}{r^4A}(\cs-1)(U-1)\left(1-r\frac{B'}{B}\right)\Bigg]+ \frac{\ca}{(\cs-1)r^2}\Bigg[2(U-2)A'f' + A^3f'{}^2\Bigg] \nonumber \\ &+\frac{2A}{r^4}\Bigg[2B^2 -\frac{r^2f'B'}{B} + r^2f''\Bigg] + \frac{8}{r^3A}(U-2)^2(U-1)\Bigg[\left(\frac{\ca}{\cs-1}\right)\frac{UA'{}^3}{A} +\frac{1}{r}\Big(1-r\frac{B'}{B}\Big)\Bigg] \nonumber \\ &+ \frac{1}{(\cs-1)r^2}\Bigg[(U-2)\left(\frac{A'{}^2}{A^3}+2f'A'\right)+A^3f'{}^2\Bigg] \Bigg[-8(\ca-\cs)(U-1) +\left(2\ca - \frac{\cs^2}{\cs-1}\right)U^2\Bigg] \nonumber \\ &+\frac{4A}{(\cs-1)r}\Bigg[\frac{2B^2(U-2)^2}{r^2} + (\cs-1)(U-2)^2f' \frac{B'}{B} + 2\cs U(U-1)\frac{f'}{r} - (\cs-1)(U-2)^2f''\Bigg]
\end{align}
\begin{align}
q_1 = \,\,&r^2\left((U-2)A'+A^3f'\right) \\
q_2 = \,\,&\Big(s_1^2 U^2 - (U-2)^2\Big)\frac{B'}{B} - \frac{2s_1^2Uq_1}{r^2A} +\frac{1}{(c_\sigma-1)A}\Big[(c_\sigma-2)U(U-2)A' +f'A^3\left(4-4c_\sigma +(c_\sigma-2)U\right)\Big]\\
q_3 = \,\,&s_1^2U\Bigg[2 (4-3U+U^2)\left(\frac{A'}{A}\right)^2 + 2AA'f'(3U-4)  + 2A^4f'{}^2 + U(U-2)\left(\frac{A'B'}{AB}-\frac{A''}{A}\right) \nonumber \\& +A^2\left(\frac{8B^2}{r^2}-Uf'\frac{B'}{B}+Uf''\right)\Bigg] + \frac{1}{(\cs-1)}\Bigg[(U-2)^2\Big(2(1-\cs)+(2-\cs)U\Big) \left(\frac{A'}{A}\right)^2 \nonumber \\ &+ 2AA'f'(U-2) \Big(2(\cs-1)+(3-2\cs)U\Big) + (\cs -2)U A^2 f'{}^2  \Bigg] \nonumber  \\ &+ (U-2)\Bigg[(U-2)^2\left(\frac{A'B'}{AB}-\frac{A''}{A}\right)+ \frac{8}{r}(U-1)\frac{A'}{A}\Bigg] +A^2\Bigg[(U-2)^2\frac{B'}{B}f' + \frac{8}{r}(U-1)f'-(U-2)^2f''\Bigg]
\end{align}
\end{widetext}

\section{Boundary conditions and the $c_\omega\to \infty$ limit}
\label{app:cw}

An elementary example that shares the most of the structure of Eqs.~\eqref{Tw} and \eqref{ae} is
\begin{gather}
c_{\omega} \omega(r)^2 + g''(r) + g(r)=0\,,\\
c_{\omega} [\omega'(r) + \omega(r)] + \sin r=0\,.\label{weq}
\end{gather}
Here, the first equation plays the role of the Einstein equations of \ae-theory [with $g''(r)+g(r)=0$  being the Einstein equations for Ho\v rava gravity], while the second represents the \ae ther equation. 
By defining $\tilde{\omega}(r)=\omega(r)\sqrt{c_\omega}$, one obtains
\begin{gather}
\tilde{ \omega}(r)^2 + g''(r) + g(r)=0\,,\\
\sqrt{c_{\omega}} [\tilde{\omega}'(r) + \tilde{\omega}(r)] + \sin r=0\,,
\end{gather}
 the general solution to which reads
\begin{align}
\tilde{\omega}(r)=&\frac{\cos r-\sin r}{2 \sqrt{c_{\omega }}}+k_1 e^{-r}\,,\\
g(r)=&\frac{e^{-2 r} }{60 c_{\omega}}\Big\{12 k_1 e^r \sqrt{c_{\omega }} (3 \sin r+\cos r)-12 k_1^2 c_{\omega }\notag\\&-5 e^{2 r} \left[-12 c_{\omega} \left(
k_3 \sin r+k_2 \cos r\right)+\sin 2 r+3\right]\Big\}\,,
\end{align}
where $k_1$, $k_2$ and $k_3$ are integration constants.
For $c_\omega\to\infty$ one then obtains
\begin{gather}
{\omega}(r)=\frac{ \tilde{\omega}(r)}{c_\omega}=\frac{k_1 e^{-r}}{\sqrt{c_\omega}}+{\cal O}\left(\frac{1}{c_\omega}\right)\,,\\
g(r)=-\frac15 k_1^2 e^{-2r}+k_2 \cos r+k_3 \sin r+{\cal O}\left(\frac{1}{\sqrt{c_\omega}}\right)\,.
\end{gather}
As can be seen, the solution for $g(r)$ in the limit $c_\omega\to\infty$ does \textit{not} satisfy equation
the ``Einstein equations'' of Ho\v rava gravity, i.e. $g''(r)+g(r)=0$. However, if we impose suitable regularity conditions on
Eq.~\eqref{weq}, e.g. such that the integration constant $k_1$ vanishes, the solution for $g(r)$ does satisfy $g''(r)+g(r)=0$. This example therefore highlights the importance of the boundary conditions for recovering (or not recovering) the Ho\v rava gravity solutions as $c_\omega\to\infty$.

\section{Regularity across multiple spin-1 horizons}
\label{app:manyspin1}

Solutions to the field equations, Eq.~\eqref{eqns:new}, are generally singular at spin-1 horizons (i.e. where $S =0$). One needs to enforce are a regularity condition --  Eq.~(\ref{eqn:reg1}) -- to avoid this from occurring. For spherically symmetric solutions with a single spin-1 horizon, this regularity condition, together with asymptotic flatness, already reduces the space of slowly-rotating solutions to a two-parameter family characterized by what could be considered the black hole's mass and spin angular momentum. Hence, when there is more than one spin-1 horizon, one can impose no further regularity conditions to keep the extra spin-1 horizons from being finite-area singularities. 

One might wish to contemplate the possibility that imposing Eq.~(\ref{eqn:reg1}) on just one spin-1 horizon (say the outermost one) ``accidentally'' renders other spin-1 horizons regular as well. Our goal here is to explicitly check that this is not the case. Note that this check cannot be done by simply generating the solutions in our setup. Our equations contain factors of $1/S$, where $S=0$ on spin-1 horizons. Such an ``accidental'' regularity would imply that each of these $1/S$ factors in the field equations should be multiplied by a quantity that vanishes just as fast as $S$ as the spin-1 horizon is approached. This is a typical $0/0$ limit that cannot be resolved numerically. Hence, we follow a different approach, which we describe below.

Let us start from the exact solution for $c_\theta = -2c_\sigma$, discussed in detail in Sec.~\ref{sec:mattingly_solns1}. This solution generally possesses two spin-1 horizons.  Recall that to obtain an asymptotically flat solution, we begin by rescaling $\Omega'$ at the outer spin-1 horizon $(r= r_{s_1})$ to 1. The regularity condition given by Eq.~\eqref{eqn:reg1} then leaves us with one parameter to tune in order to find an asymptotically flat solution. In practice, this parameter is $\Lambda(r_{s_1})$. By bracketing/shooting, a unique value for $\Lambda(r_{s_1})$ is found that gives an asymptotically flat solution. Let us call this solution $\{\Lambda_1(r),\Omega'_1(r)\}$.

We can attempt to match this asymptotically flat solution to another solution that is regular at the inner spin-1 horizon ($r=r_{s_2}, r_{s_2} < r_{s_1}$). Again we set $\Omega'(r_{s_2}) =1$, and by imposing Eq.~\eqref{eqn:reg1} at $r=r_{s_2}$ we are only left with one parameter, $\Lambda(r_{s_2})$, to specify. For any choice of $\Lambda(r_{s_2})$, one can integrate outward and get a corresponding solution. Let us call this $\{\Lambda_2(r),\Omega'_2(r)\}$.

To determine if the regularity conditions at both spin-1 horizons can be simultaneously satisfied, we check if any of the solutions $\{\Lambda_2(r),\Omega'_2(r)\}$, which are regular at the inner spin-1 horizon (and depend on $\Lambda(r_{s_2})$ as input), are linearly related to the asymptotically-flat solution $\{\Lambda_1(r),\Omega'_1(r)\}$. We do this by looking at the Wronskian of pairs of solutions at the midpoint $r_m = (r_{s_1}+r_{s_2})/2$, i.e. we compute the following quantities at $r= r_m$: 
\begin{align}
\mbox{w}_\Lambda &:= \frac{\Lambda'_2}{\Lambda_2} - \frac{\Lambda'_1}{\Lambda_1} = \frac{W[\Lambda_1,\Lambda_2]}{\Lambda_1\Lambda_2},  \\
\mbox{w}_\Omega &:= \frac{\Omega''_2}{\Omega'_2} - \frac{\Omega''_1}{\Omega'_1} = \frac{W[\Omega'_1,\Omega'_2]}{\Omega'_1\Omega'_2},  
\end{align}
where $W[f_1,f_2] := f_1 f'_2 - f_2 f'_1$ is the Wronskian of $\{f_1,f_2\}$, and evaluate 
\begin{align}
\Delta:= \sqrt{\mbox{w}_\Lambda^2 + \mbox{w}_\Omega^2}. 	
\end{align}
That $\Delta$ vanishes at $r_m$ is a necessary (and sufficient) condition for smoothly joining the two solutions. We thus scan the parameter space for $\Lambda(r_{s_2})$ seeking a value that results in $\Delta = 0$ (to within our numerical accuracy). We have not succeeded in finding such a value, and our numerical results suggest that it may not exist.

One point of contention about this test is that the rescaling that sets $\Omega'(r_{s_2}) =1$ is not the same as the one that gives $\Omega'(r_{s_1}) =1$. We note, however, that $\mbox{w}_\Lambda$ and $\mbox{w}_\Omega$ (and thus $\Delta$) are invariant under such rescalings (i.e. $\{\Lambda(r),\Omega'(r)\} \rightarrow \{K\Lambda(r),K\Omega'(r)\}$). Therefore, the test above does not depend on what we choose for $\Omega'(r_{s_2})$, and setting $\Omega'(r_{s_1}) =1$ and $\Omega'(r_{s_2}) =1$ at the same time is justified.

We performed similar tests for solutions to the field equations in the small-coupling limit where we find multiple spin-1 horizons. Again, we first determine the unique asymptotically-flat solution to Eq.~\eqref{eqn:ueqnother} that is regular at the outermost spin-1 horizon. We then integrate this solution inward from the outermost spin-1 horizon, at $r= r_{s_1}$, down to the next spin-1 horizon, $r= r_{s_2}$. Let us call this solution $\Lambda_1(r)$.  
We then derive a second solution (which we call $\Lambda_2(r)$)
by imposing regularity at $r= r_{s_2}$. This solution is completely determined once the value at $r= r_{s_2}$,  $\Lambda(r_{s_2})$, is specified.

To see if we can smoothly join $\Lambda_1(r)$ and $\Lambda_2(r)$, we then scan the parameter space for $\Lambda(r_{s_2})$ and compute 
\begin{equation}
\sqrt{(\Lambda_1(r_m)-\Lambda_2(r_m))^2+(\Lambda_1'(r_m)-\Lambda_2'(r_m))^2}
\end{equation}
in search of possible zeroes. Again, we have not found any such zeros, which implies that the two solutions cannot be matched smoothly.
As a technical side point, we recall that when deriving the field equations in the small-coupling limit, we set (without loss of generality) $\kappa=1$ in $\Omega'(r) = \kappa/r^4$. This is indeed required to arrive at Eq.~\eqref{eqn:ueqnother}. This is analogous to what we do above, when we use the homogeneity of the field equations in the
slow-rotation limit to set $\Omega'(r_{s_1}) =1$ and $\Omega'(r_{s_2}) =1$.

In summary, our numerical results suggest that even when regularity is imposed at the outermost spin-1 horizon, the other spin-1 horizons will generically be singular.

\bibliography{myrefs}

\end{document}